\documentclass[%
reprint,
superscriptaddress,
nofootinbib,
 amsmath,amssymb,
 aps,
pre
]{revtex4-2}

\usepackage[utf8]{inputenc}
\usepackage[english]{babel}
\usepackage{graphicx}
\usepackage{dcolumn}
\usepackage{bm}
\usepackage{mathtools}

\usepackage{xcolor}

\usepackage{dcolumn}
\usepackage{xspace} 

\usepackage{xcolor}
\definecolor{darkgreen}{rgb}{0,0.45,0}
\usepackage[unicode,
  colorlinks = true,
            linkcolor = blue,
            urlcolor  = blue,
            citecolor = blue,
            anchorcolor = blue,
    filecolor=blue]{hyperref}

\usepackage{tikz}

\usepackage{graphicx}

\usetikzlibrary{decorations.pathmorphing}
\usetikzlibrary{decorations.markings}
\usetikzlibrary{arrows,shapes,snakes,automata,backgrounds,petri}
\usetikzlibrary{calc}
\usetikzlibrary{patterns}
\usetikzlibrary{decorations.text}

\tikzset{
xxtsubstrate/.style={decorate, 
line width=1pt,
draw=olive, 
decoration=snake, 
segment amplitude=0.75mm, 
line after snake=0.25mm,
line before snake=0.25mm
},
tsubstrate/.style={decorate, 
line width=1pt,
draw=olive, 
decoration=snake, 
segment amplitude=0.5mm, 
segment length=5pt,
segment amplitude=0.2mm, 
line after snake=1mm,
line before snake=1mm
},
Bsubstrate/.style={decorate, 
line width=1pt,
draw=darkgreen, 
decoration=snake,
segment length=5pt,
segment aspect=0,
segment amplitude=0.5mm, 
line after snake=0mm,
line before snake=0mm
},
substrate/.style={decorate, 
line width=1pt,
draw=green, 
decoration=snake, 
segment length=5pt,
segment amplitude=0.5mm, 
line after snake=0.5mm,
line before snake=0.5mm
},
activity/.style={very thick,draw=red,postaction={decorate},
decoration={markings,mark=at position .5 with
{\arrow[draw=red]{>}}}},
tactivity/.style={thick,draw=red,postaction={decorate},
decoration={markings,mark=at position .5 with
{\arrow[draw=red]{>}}}},
tEPSactivity/.style={thick,draw=red,postaction={decorate},
decoration={markings,mark=at position .55 with
{\arrow[draw=red]{>}}}},
tAactivity/.style={thick,draw=red},
Aactivity/.style={very thick,draw=red},
Bactivity/.style={very thick,draw=blue,dashed},
Cnoise/.style={very thick,draw=blue,dashed},
tSactivity/.style={thick,draw=red,postaction={decorate},
decoration={markings,mark=at position .7 with
{\arrow[draw=red]{>}}}},
Sactivity/.style={very thick,draw=red,postaction={decorate},
decoration={markings,mark=at position .7 with
{\arrow[draw=red]{>}}}}
}

\newcommand{\Eqref}[1]{Eq.~\eqref{#1}}
\newcommand{\Eqrefs}[1]{Eqs.~\eqref{#1}}

\def\LL{\mathcal{L}}
\def\adj{^{\dagger}}
\newcommand{\dd}[1]{{\rm d}{#1}\,}
\newcommand{\dint}[1]{{\rm d}{#1}\,}
\newcommand{\dbar}{{\rm d}\hspace*{-0.08em}\bar{}\hspace*{0.1em}}
\newcommand{\ddbar}{{\rm d}\hspace*{-0.08em}\bar{}\hspace*{0.1em}^d}
\newcommand{\deltabar}{{\delta}\hspace*{-0.08em}\bar{}\hspace*{0.1em}}
\let\upphi\phi
\let\phi\varphi

\def\del{\partial}
\def\eps{\varepsilon}

\def\bP{\mathbb{P}}

\def\cD{\mathcal{D}}

\def\cI{\mathcal{I}}
\def\cJ{\mathcal{J}}

\def\cO{\mathcal{O}}
\def\cP{\mathcal{P}}
\def\cQ{\mathcal{Q}}

\def\cS{\mathcal{S}}

\def\cf{cf.~}
\def\ie{i.e.,~}
\def\eg{e.g.,~}

\renewcommand{\iint}{\int \!\!\!\!\! \int}

\def\target{x}
\def\start{x_0}

\def\deprate{\gamma}
\def\latt{\delta_a}

\def\del{\partial}

\def\tphi{\widetilde{\varphi}}
\def\tpsi{\widetilde{\psi}}

\def\cpl{g}
\def\Smob{\cS_{\phi}}
\def\Strc{\cS_{\psi}}
\def\Sdep{\cS_{\gamma}}

\def\trace{\cQ}
\def\traceI{\cQ^{(1)}_{\mathrm{I}}}
\def\traceII{\cQ^{(1)}_{\mathrm{II}}}
\def\traceIII{\cQ^{(1)}_{\mathrm{III}}}
\def\traceIV{\cQ^{(1)}_{\mathrm{IV}}}

\def\fptxy{\tau_{x_0,x_1}}
\def\fptdist{\bP_{\text{FPT}}}
\def\fptmgf{\chi_{\text{FPT}}}

\def\maxdist{\bP_{\text{Max}}}

\def\vol{\left\langle\operatorname{Vol}\right\rangle}

\def\tx{\mathfrak{x}}

\newcommand{\transm}[1]{T\left({#1}\right)}
\newcommand{\returnm}[1]{R\left({#1}\right)}

\newcommand{\transeff}[1]{T^{(2)}\left({#1}\right)}
\newcommand{\returneff}[1]{R^{(2)}\left({#1}\right)}

\newcommand{\avg}[1]{\left\langle#1\right\rangle}
\newcommand{\yavg}[1]{\overline{#1}}
\newcommand{\corr}[1]{C_2(#1)}
\newcommand{\fcorr}[1]{\hat{C}_2(#1)}
\newcommand{\invcorr}[1]{\bar{C}_2(#1)}
\newcommand{\trans}[1]{{T}\left(#1\right)}
\newcommand{\return}[1]{{R}\left(#1\right)}

\newcommand{\party}{\mathcal{Z}_y}
\newcommand{\tw}{\tilde{\omega}} 

\newcommand{\carryingCap}{n_0}

\usepackage[normalem]{ulem}
\usepackage{subfigure}
\newcommand{\new}[1]{#1}

\begin{document}

\title{Field theory of survival probabilities, extreme values, first passage times, and mean span of non-Markovian stochastic processes}

\author{Benjamin Walter}
\affiliation{Department of Mathematics, Imperial College London, 180 Queen's Gate, SW7 2AZ London, United Kingdom}
\affiliation{Centre for Complexity \& Networks, Imperial College London, SW7 2AZ London, United Kingdom}
\affiliation{SISSA-International School for Advanced Studies, via Bonomea 265, 34136 Trieste, Italy}
\affiliation{INFN, Sezione di Trieste, 34136 Trieste, Italy}
\email{bwalter@sissa.it}

\author{Gunnar Pruessner}
\affiliation{Department of Mathematics, Imperial College London, 180 Queen's Gate, SW7 2AZ London, United Kingdom}
\affiliation{Centre for Complexity \& Networks, Imperial College London, SW7 2AZ London, United Kingdom}

\author{ Guillaume Salbreux}
\affiliation{The Francis Crick Institute, 1 Midland Road, NW1 1AT London, United Kingdom}
\affiliation{Department of Genetics and Evolution, University of Geneva, Quai Ernest-Ansermet 30, 1205 Geneva, Switzerland}
\date{\today}

\date{\today}

\begin{abstract}
We provide a perturbative framework to calculate extreme events of non-Markovian processes, by mapping the stochastic process to a two-species reaction diffusion process in a Doi-Peliti field theory combined with the Martin-Siggia-Rose formalism. 
This field theory treats interactions and the effect of external, possibly self-correlated noise in a perturbation about a Markovian process, thereby providing a systematic, diagrammatic approach to extreme events.
We apply the formalism to Brownian Motion and calculate its survival probability distribution subject to self-correlated noise. 
\end{abstract}

\maketitle

\section{Introduction \label{sec:intro}}
Many non-equilibrium systems are studied by projecting out a single slow degree of freedom which evolves stochastically and often displays non-negligible memory effects \cite{Zwanzig1960,Haake1973,zwanzig2001}.  
A classic object of study is its survival probability which describes the probability of the degree of freedom not having reached a  threshold yet
\cite{Majumdar1999,bray_persistence_2013}. 
The survival probability defines not only the persistence exponents \cite{Aurzada2015}, but is also closely linked to the distribution of first-passage times, running maxima, and spans via some simple relations. 
All of these \emph{extreme events}  aptly characterise the non-equilibrium nature of complex systems and have been studied separately over the last hundred years (for classic references see \cite{erwin_schrodinger_zur_1915,daniels_smithies_1941,gnedenko_1943,darling_first_1953,Gumbel_1958}, for recent overviews \cite{metzler_first-passage_2014,majumdar_extreme_2020}).
 
Survival probabilities of non-Markovian processes are, however, notoriously hard to compute as they depend on the entire trajectory whose distribution is usually impossible to obtain \cite{prigogine_colored_2007,Wiese2011,majumdar_extreme_2020,singh_extremal_2021}.
Perturbative schemes such as  \cite{Majumdar1996,Oerding1997} have proven to be successful in characterising the behaviour of the survival probabilities for large times in classical non-equilibrium models. 
More recently, a similar perturbation theory has been applied to fractional Brownian Motion to further access the full survival probability in a perturbation theory about the Hurst parameter \cite{Wiese2011,Arutkin2020}. These  techniques, however, heavily rely on Gaussianity and do not readily translate to more general non-Gaussian non-Markovian processes.

In this article, we compute the \emph{full} survival probability of processes subject to both uncorrelated and self-correlated noise in a perturbation theory in the strength of the self-correlated noise. 
A physical realisation of such processes is a particle  immersed in a heat bath and subject to a random self-propelling force. 
Our perturbative framework is valid in the regime where the self-correlated noise is small compared to thermal fluctuations stemming from the heat bath.
More generally,
this type of process is central to the study of active matter \cite{malakar_steady_2018,sevilla_generalized_2019,dabelow_irreversibility_2019,Shee2020,Martin2021} and  non-equilibrium phenomena \cite{luczka_non-markovian_2005,caprini_active_2018,caprini_entropy_2019,caprini_entropy_2019}.

In our recent work \cite{Walter2021}, we presented a scheme to calculate first-passage distributions for the same class of non-Markovian processes. These results relied on a perturbative functional expansion of a renewal type equation inspired by the classical work of \cite{siegert_first_1951,darling_first_1953}. 
Using a field-theoretic approach that draws on both the Doi-Peliti  \cite{Doi1976,Peliti1985} as well as the Martin-Siggia-Rose formalism \cite{Martin1973},
in the present work we generalise these results to a broader class of extreme events.

Using a field theory has some notable advantages. Firstly, the diagrammatics give a clear intuition of the underlying microscopic processes otherwise hidden within cumbersome expressions. 
Secondly, the field theory provides a systematic perturbative framework naturally drawing on renormalisation techniques.
Thirdly, it is easily extended to incorporate further interactions, such as reactions, external and pair potentials.

This article is structured as follows. First, we map a Markovian process onto a field theory. Secondly, we introduce a field-theoretic mechanism which is designed to keep track of the space already visited by the process. This defines the visit probability $\trace(\start, \target, t)$, the probability that the process started at $\start$ has been at $\target$ prior to time $t$, and which is the complement of the survival probability. Thirdly, we add the self-correlated driving noise, thus breaking Markovianity, and compute the  corrections induced in $\trace$. Finally, we illustrate the approach by computing the correction to the survival probability of Brownian Motion driven by self-correlated noise.

\section{Field theory for Markovian visit probabilities \label{sec:ft_markov}}

\subsection{Markovian transition probabilities}

In this article, we construct a perturbation theory around Markovian processes  characterised by a Langevin Equation \cite{kampen_stochastic_2007},
\begin{align}
	\nonumber
	\dot{x}_t &= -V^{\prime}(x_t) + \xi_t \\
	x(t=t_0)&=x_0 
	\label{eq:langevin}
\end{align}
where $V^{\prime}(x_t)$ is the gradient of a potential and $\xi_t$ Gaussian white noise with correlator $\avg{\xi_t \xi_{t'}} = 2D_x \delta(t-t')$. Further, we introduce $T(x,t) \equiv T(x_0,x;t_0,t)$ as the \emph{transition probability} for the walker to travel from $x_0$ at time $t_0$ to $x$ at time $t$.
This probability density is also known as Green's function or propagator in related fields of mathematics.
We will state $x_0$ and $t_0$ only where needed for clarity.
\new{The transition probability satisfies a Fokker-Planck equation \cite{risken_fokker-planck_1984}}
\begin{align}
\partial_t T(x,t) = \left(V^{\prime \prime}(x) + V^{\prime}(x) \partial_x + D_x \partial_x^2 \right) T(x,t) 
\label{eq:fokker_planck}
\end{align}
\new{with initial condition $T(x,t_0) = \delta(x-x_0)$.} 

As is detailed in \cite{Garcia-MillanPruessner:2021}, the process \eqref{eq:langevin} can be mapped to a Doi-Peliti field theory \cite{TaeuberHowardVollmayr-Lee,Cardy:2008,Garcia-MillanPruessner:2021} containing two fields, the annihilator field $\phi(x,t)$ and the creator field $\phi^\dagger(x,t)= 1+ \tphi(x,t)$, which are jointly distributed according to
\begin{align}
\cP[\phi,\tphi] = \exp\left(-\Smob[\phi,\tphi]\right) \ .
\label{eq:smob_probability}
\end{align}
Here the action $\Smob[\phi,\tphi]$ is constructed as
\begin{multline}
    \Smob 
    = \iint \dd{x}  \dd{t}   \tilde{\varphi}\left(\partial_t -  V^{\prime\prime}(x) - V^\prime(x)\partial_x - D\partial_x^2 \right) \varphi.
    \label{eq:smob_def}
\end{multline}
Moreover, the transition probability satisfies
\begin{align}
	T(x,t) = \avg{\phi(x,t)(1+\tphi(x_0,t_0))}_{\Smob}, \label{eq:trans_as_avg}
\end{align}
where $\avg{\bullet}_{\Smob}$ denotes the expectation over the measure \eqref{eq:smob_probability}.

\new{Constructing a solution to the partial differential equation in \Eqref{eq:fokker_planck} via a path integral can in principle be done with the Feynman-Kac theorem \cite{Kac1949,ito1961}. Here, however, we use a non-equilibrium field theory following \cite{Martin1973} (see Sec.~\ref{sec:discussion_qft} for further discussion).}

\subsection{Visit probability and extreme events \label{sec:visit_prob}}
\begin{figure}[h!]
\includegraphics[width=\columnwidth]{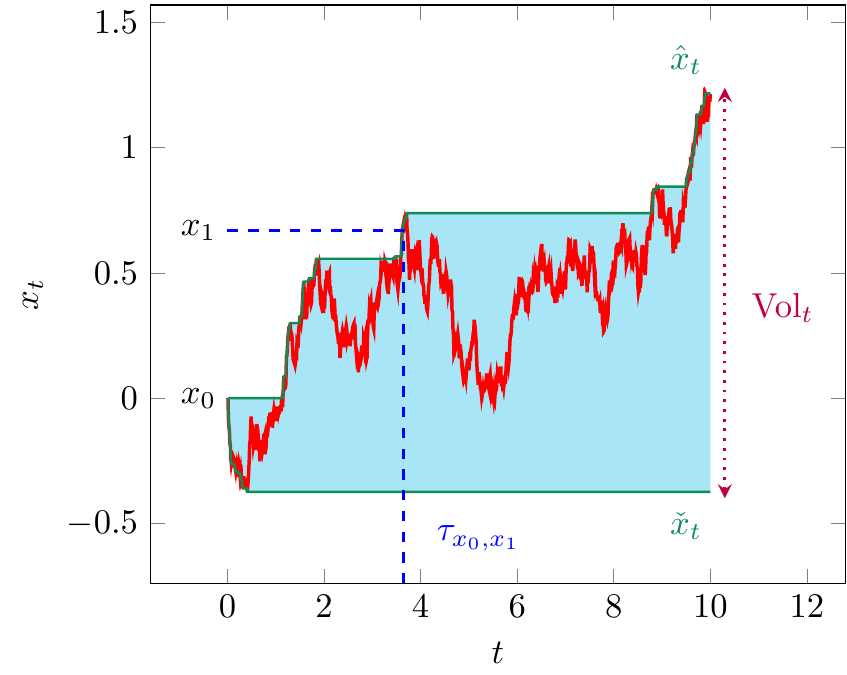}
\caption{A continuous random path $x_t$ (red line) evolves in time. The \emph{visited area} is shaded in blue. In this article, we study the visit probability $\trace(x,t)$, \ie the probability that a point $(x,t)$ lies within the blue area. As discussed in Sec.~\ref{sec:visit_prob}, the visit probability allows one to compute the distribution of \emph{(i)} first-passage times $\tau_{x_0,x_1}$ (blue dashed line, \Eqref{eq:def_fptdist}) and \emph{(ii)} running maxima $\hat{x}_t$ (green line, \Eqref{eq:def_maxdist}), as well as \emph{(iii)} the average volume explored (purple arrow, \Eqref{eq:mean_vol}).}
\label{fig:Q_illustration}
\end{figure}

The key problem we address here is how to approximate the distribution of first-passage times, running maxima, and mean volume explored of the process defined in \Eqref{eq:langevin}. These \emph{extreme events} are all mutually related via the \emph{visit probability} which we define as 
$
	\trace(x_0,x,t_0,t) = \trace(x,t) 
	= \bP [ x_s = x \text{ at some time } t_0 \leq s \leq t],$
\ie the complement of the \emph{survival probability} $\mathbb{P}_{\text{surv}} = 1 - \trace$. This measures the probability that the particle has been at $x$ at or before time $t$.
In Fig.~\ref{fig:Q_illustration}, we show a single realisation of $x_t$, together with its visited area.

The visit probability contains various informations about the process: When taking the derivative $\partial_{t}\trace(x,t)$, one measures the weight of those paths which visit $x$ at $t$ for the first time. The latter is the first-passage time, shown in Fig.~\ref{fig:Q_illustration} as a blue dashed line, and thus its distribution satisfies 
\begin{align}\label{eq:def_fptdist}
\fptdist (\fptxy = t) = \del_t \trace(x_1,t).
\end{align}
Analogously, taking the derivative  $-\partial_{x}\trace(x,t)$, weighs those paths who at time $t$ visit $\target$ for the first time, or alternatively, the distribution of the maximum $\hat{x}_t = \max_{s\leq t} x_s$, shown as a green line in Fig.~\ref{fig:Q_illustration}, \ie 
\begin{align}\label{eq:def_maxdist}
\maxdist(\hat{x}_t = x) = -\partial_{x}\trace(x,t),
\end{align} 
\new{for $ x >x_0$.}
Moreover, integrating over $\int \dint{x} \trace(x,t)$, gives the average of the volume explored, which is defined as the difference between running maximum and minimum, $\operatorname{Vol}([x_t],t) = \max_{s\leq t} x_s - \min_{s \leq t} x_s$, as illustrated in Fig.~\ref{fig:Q_illustration}. Its mean is given by 
\begin{align}
    \vol = \int \dint{x} \trace(x,t).
    \label{eq:mean_vol}
\end{align}
Higher moments of the volume explored are considered in \cite{bordeu_volume_2019}.

\subsection{\new{Overview of main results}}
\new{In the following, we build a framework to compute the visit probability $\trace(x,t)$ for a specific class of non-Markovian processes. These are given by the solution to the stochastic differential equation
\begin{align}
\dot{x}_t = - V^{\prime}(x_t) + \xi_t + \cpl y_t\ .
\label{eq:driven_lgv}
\end{align} 
\new{This equation extends the Markovian Langevin equation} 
\eqref{eq:langevin} by adding a second independent noise term $y_t$ which is assumed to be stationary, of zero mean, but not necessarily Gaussian. 
The driving noise $y_t$  further carries dimensions of a velocity, leaving $\cpl$ as a dimensionless coupling constant which we suppose to be small. The main result of this article then is a perturbative expansion of the visit probability of $x_t$ to leading order in $\cpl^2$, \ie assuming the visit probability allows for an analytical expansion around $g=0$ as
$\trace(x,t) = \trace^{(0)}(x,t) +\cpl^2 \trace^{(2)}(x,t) + \cO(\cpl^3)$, we find formulas for the correction terms.  
As is further detailed below, the assumption that $\trace(x,t)$ be analytic in $\cpl$ restricts the possible choice of driving noises depending on the choice of potential $V(x_t)$.
Together,  Eq.~\eqref{eq:nonM_trace} and \Eqref{eq:T2_formula} provide a general formula for the leading perturbative correction, 
$\trace^{(2)}(x,t)$
, which is expressed in terms of the Markovian transition probability $T(x,t)$ and the two-time  correlation function of 
$y_t$, 
\begin{align}
C_2(t-s) = \overline{y_s y_t}
,
\end{align}
 where 
$\overline{\bullet}$ 
denotes the average with respect to the path measure of $y_t$. }

In principle, the framework also allows to compute higher-order corrections, \ie $g^n \trace^{(n)}(x,t)$, using the $n$ point correlations of the driving noise. In the presentation of the results, however, we restrict ourselves to the leading order perturbation only.

The processes described by \Eqref{eq:driven_lgv} do not satisfy a (generalised) fluctuation-dissipation relation, and hence cannot be brought into the form of a generalised Langevin Equation. Instead, these processes are often used to model active matter in thermal environments \cite{malakar_steady_2018,sevilla_generalized_2019,dabelow_irreversibility_2019,Shee2020,Martin2021,luczka_non-markovian_2005,caprini_active_2018,caprini_entropy_2019}, which typically operate away from equilibrium.

Finally, although the expression for the visit probability applies to all potentials $V(x)$, and can be employed numerically to study these, an analytically closed expression can only be expected in cases where an analytic solution to the Markovian Fokker Planck equation \eqref{eq:fokker_planck} is known. This effectively reduces the class of potentials for which we obtain analytical results to harmonic or flat potentials, \ie perturbations of Brownian Motion or Ornstein-Uhlenbeck processes.

\subsection{Markovian visit probabilities}

In this section, we present a field theory of visit probabilities for Markovian processes. Whilst in the case of transition probabilities it is well known that the solution to the Fokker Planck equation \eqref{eq:fokker_planck} can be expressed as a path integral, \Eqref{eq:trans_as_avg}, this has not yet been established for the visit probability 
$\trace(x,t)$. Our aim is to construct a field theory whose correlation functions equal 
$\trace(x,t)$, in close analogy to \Eqref{eq:trans_as_avg}. In difference to the case for the transition probability, however, this field theory cannot be straightforwardly constructed for 
$\trace(x,t)$, since no evolution equation for 
$\trace(x,t)$, comparable to \Eqref{eq:fokker_planck}, exists to our knowledge.

As is explained in great detail in \cite{nekovar_field-theoretic_2016,bordeu_volume_2019,amarteifio_2019}, and briefly discussed in App.~\ref{app:tracing_mechanism}, the visit probability $\trace(x,t)$ can be expressed as a field-theoretic expectation value under a  Doi-Peliti field theory by introducing two additional auxiliary (``trace''-) fields $\psi(x,t)$ and $\tpsi(x,t)$ with a joint distribution
\begin{multline}
	\cP[\phi,\tphi,\psi,\tpsi] \\
	= \lim_{\gamma \to \infty}\exp \left(- \Smob[\phi,\tphi] -\Strc[\psi,\tpsi] + \gamma \Sdep[\phi,\tphi,\psi,\tpsi]  \right)
	\label{eq:joint_field_measure}
\end{multline}
such that the visit probability can be written as
\begin{align}
	\trace(x,t) =  \new{\carryingCap^{-1}}
 \avg{\psi(x,t)\left(1+ \tphi(x_0,t_0) \right)}_{\cS}
	\label{eq:trace_as_avg}
\end{align}
where $\avg{\bullet}_{\cS}$ is understood as the average with respect to the measure in \Eqref{eq:joint_field_measure}. \new{Here, we have introduced a normalising density $\carryingCap$ which is further detailed below and in App.~\ref{app:tracing_mechanism}.}

\new{The pair of fields $\psi, \tpsi$ are a stochastic auxiliary variable which tracks the volume explored by the process $x_t$ (see Fig.~\ref{fig:Q_illustration}). This is in analogy to $\phi, \tphi$ whose correlation tracks the current position of the process $x_t$ (\cf \Eqref{eq:trans_as_avg}). Hence, measuring the average field density $\psi(x,t)$, and normalising by a unit density $\carryingCap$, amounts to computing the probability that the process visited $x$ up to time $t$. The average in \Eqref{eq:trace_as_avg} then corresponds to the probability density that $x$ has been visited prior to $t$ \emph{conditioned} on the process having been initialised at $x_0$ at time $t_0$; thus matching our definition of the visit probability.}

The field action $\Smob + \Strc - \gamma \Sdep$ consists of three actions which model $(i)$ the diffusion of the process $x_t$ (\cf \Eqref{eq:smob_def}), $(ii)$  the non-interacting dynamics of the auxiliary fields tracking the volume explored by the particle, and $(iii)$ the interaction between the random process $x_t$ and the auxiliary fields tracking its explored volume. Clearly, the explored volume depends on all the previous positions of the process $x_t$ and therefore the third contribution contains all four fields $\phi, \tphi ,\psi, \tpsi$. It is multiplied with a rate $\gamma$ to be taken to $\infty$. This rate is interpreted as the rate with which the process $x_t$ ``traces'', \ie marks as explored, a given point $x$. Taking $\gamma \to \infty$ amounts to  the field $\psi$ tracking \emph{every} visited point.\footnote{Leaving $\gamma$ finite amounts to imperfect tracking, suitable to study imperfect reaction kinetics  as discussed in, \eg \cite{Guerin2021}. } In what follows, we outline the components of the action in \Eqref{eq:joint_field_measure} and refer the reader to the appendices for technical details. 

The first term in the exponential of \Eqref{eq:joint_field_measure} is the \emph{diffusion action} $\Smob$ introduced in \eqref{eq:smob_def}.
The second term $\Strc$ denotes the \emph{trace action}
\begin{align}
	 \Strc  = \iint \dd{x} \dd{t} \tpsi(\partial_t + \varepsilon)\psi \ .
	\label{eq:trace_action_def}
\end{align}
Comparing with $\Smob$, \Eqref{eq:smob_def}, the trace action can be interpreted as corresponding to the process \eqref{eq:langevin} with $V(x) \equiv 0, D_x = 0$, \ie a deterministic immobile particle. This reflects  the fact that a point, once visited by the process $x_t$, remains \emph{visited} forever. To ensure convergence of the path integral over \eqref{eq:joint_field_measure}, we have included a positive parameter $\eps > 0$. Latter further ensures causality \cite{cardy1999,bordeu_volume_2019}, \ie that a point is only marked visited \emph{after} it has been visited by the process $x_t$. In a field-theoretic context, the parameter $\eps$ is referred to as a \emph{mass} or an \emph{infrared regulator} as it suppresses the divergencies otherwise arising in \Eqref{eq:joint_field_measure} from the contributions of \Eqref{eq:trace_action_def} at large times. The parameter $\eps$ is to be taken to zero at the end of the calculation. 

The third term of the exponential of \Eqref{eq:joint_field_measure} is the
\emph{deposition action}. It describes the growth of the volume explored due to fluctuations of the process $x_t$ and is derived in \cite{nekovar_field-theoretic_2016}. In the continuum limit, it reads
\begin{align}
	\Sdep =  \iint \dd{x} \dd{t} \left( \tau \tpsi\varphi + \sigma \tphi\tpsi\varphi - \lambda \tpsi\varphi\psi - \kappa \tphi\tpsi\varphi \psi \right) \ .
	\label{eq:sdep_def}
\end{align}

The four dimensionfull couplings $\tau$, $\sigma$, $\kappa$ and $\lambda$, are introduced as differently for independent renormalisation.
However, the rates $\tau$ and $\sigma$ and the densities $\kappa, \lambda$ are each equal  and are related to each other via 
\begin{align}
\lambda = \kappa = \carryingCap^{-1} \tau = \carryingCap^{-1} \sigma ,
\label{eq:bare_couplings}
\end{align} 
 see App.~\ref{app:tracing_mechanism} for details. Overall, $\Sdep$ is multiplied with a dimensionless constant $\gamma$ which needs to be taken to $\gamma \to \infty$.

Each of the four vertices in \Eqref{eq:sdep_def} is diagrammatically represented as
\begin{align}\label{eq:vertices}
    \tikz[baseline=-2.5pt]{
    \node at (0,.3) {$\tau$};
    \draw[Aactivity] (0,0) -- (.5,0) ;
    \draw[Bsubstrate] (-.5,0) -- (0,0); 
    }\qquad
     \tikz[baseline=-2.5pt]{
      \node at (0,.3) {$\sigma$};
    \draw[Aactivity] (-.5,0) -- (.5,0) ;
    \draw[Bsubstrate] (0,0) -- (-.5,-.4);
    }\qquad
    \tikz[baseline=-2.5pt]{
     \node at (0,.3) {$-\lambda$};
    \draw[Aactivity] (0,0) -- (.5,0) ;
    \draw[Bsubstrate] (-.5,0) -- (0,0);
    \draw[Bsubstrate] (0,0) -- (.5,-.4);
    } \qquad 
      \tikz[baseline=-2.5pt]{
       \node at (0,.3) {$-\kappa$};
    \draw[Aactivity] (-.5,0) -- (.5,0) ;
    \draw[Bsubstrate] (-.5,-.4) -- (0,0);
    \draw[Bsubstrate] (0,0) -- (.5,-.4);}
\end{align}
and enters into the action multiplied by $\gamma$\new{, see \Eqref{eq:joint_field_measure}}. Here, straight red lines represent the propagators of $\phi, \tphi$, and green wriggly lines those of $\tpsi, \psi$. As a convention, we read vertices/diagrams  from right to left.

It follows that the diagrammatic expansion of  the trace function \eqref{eq:trace_as_avg} is
\begin{align}
	\trace(x,t) = \carryingCap^{-1} \lim_{\gamma \to \infty}
    \tikz[baseline=-2.5pt]{
    \draw[Aactivity] (0,0) -- (.7,0) node [at end, above] {$(x_0, t_0)$}; 
    \draw[Bsubstrate] (-.7,0) -- (.,0) node [at start, above] {$(x,t)$};
    \fill (0,0) circle(3pt);}
\end{align}
where the central dot stands for the renormalised coupling $ \tau_R$ \new{ and the limit in $\gamma \to \infty$ stems from the definition of the action  in \Eqref{eq:joint_field_measure}}.  This renormalisation is given by the diagrammatic expansion
\begin{multline}
     \tau_R 
     = \tikz[baseline=-2.5pt]{
    \draw[Aactivity] (0,0) -- (.3,0) ; 
    \draw[Bsubstrate] (-.3,0) -- (.,0) ;
    \fill (0,0) circle(3pt);}
    \\
    = \gamma \tikz[baseline=-2.5pt]{
    \node at (0,.3) {$\tau$};
    \draw[Aactivity] (0,0) -- (.3,0) ;
    \draw[Bsubstrate] (-.3,0) -- (0,0); 
    }
    + \gamma^2
    \tikz[baseline=-2.5pt]{
    \node at (0,.3) {$-\lambda$};
    \node at (0.5,.3) {$\sigma$};
    \draw[Bsubstrate] (-0.3,0) -- (-0,0) ;
    \draw[Aactivity] (0,0) -- (.8,0) ;
    \draw[Bsubstrate] (0.0,0) arc (180:360:.25);
    }
     + \gamma^3
    \tikz[baseline=-2.5pt]{
        \node at (0,.3) {$-\lambda$};
    \node at (0.5,.3) {$-\kappa$};
    \node at (1.0,.3) {$\sigma$};
    \draw[Bsubstrate] (-0.3,0) -- (-0,0) ;
    \draw[Aactivity] (0,0) -- (1.3,0) ;
    \draw[Bsubstrate] (0.0,0) arc (180:360:.25);
    \draw[Bsubstrate] (0.5,0) arc (180:360:.25);
    }+ ...
    \label{eq:tau_r_diagram_sum} 
    \end{multline}
The only diagrams contributing to this expansions are chains of the loop-diagram $\tikz[baseline=-5pt]{
    \draw[Aactivity] (0,0) -- (.5,0) ;
    \draw[Bsubstrate] (0.0,0) arc (180:360:.25);}$.
\new{We introduce the \emph{return probability} $\return{x,t} = \trans{x,x,t}$ and likewise its Fourier transform $\return{x,\omega}$.}
As shown in App.~\ref{app:markovian_visit_p} 
the fully renormalised vertex can be evaluated using a geometric  sum  as
    \begin{align}
	    \tau_R(\omega) 
	    = \frac{\gamma\tau}{1+\gamma\kappa\return{x,\omega}}
	    \label{eq:eff_tau_coupling} 
    \end{align}
    which is an \emph{exact} result for all $\gamma$, 
    such that the effective trace function, Fourier transformed, is
    \begin{multline}
    \label{eq:Q_in_omega}
	    \int \dint{t}e^{i\omega t} \trace(x,t)  
	    =   \carryingCap^{-1} \frac{1}{-i\omega + \eps}  \tau_R(\omega)\trans{\start,\target,\omega}  \\
	    = \frac{1}{-i\omega } \frac{\trans{\start,\target,\omega} }{ \return{\target,\omega}}    \end{multline}
    where we made use of time-translational invariance to write the Fourier-transform in one frequency only, tacitly took the limit $\eps \to 0$, and used \new{$\tau/\kappa = \carryingCap$, see \Eqref{eq:bare_couplings}}. Multiplying this result with $(-i\omega)$ gives the first-passage time moment generating function. Then, the result in \eqref{eq:Q_in_omega} agrees with classical results given in \cite{siegert_first_1951,darling_first_1953}.

\new{In summary, by introducing an extended field theory via the four-field action in \Eqref{eq:joint_field_measure}, the visit probability of a Markovian process can be written as the field-theoretic average \eqref{eq:trace_as_avg}, in direct analogy to the (simpler) transition probability which can be represented with two fields, \cf \Eqref{eq:trans_as_avg}. This effective field theory for visit probabilities comes at the cost of the additional fields $\psi, \tpsi$ which need to be related to the fields $\phi, \tphi$ by a nonlinear interaction $\Sdep$. The effect of this interaction on the growth of the volume explored by $x_t$ can be captured by evaluating the effective (``renormalised'') deposition vertex $\tau_R$. Formally, this amounts to evaluating an infinite series of correction terms, diagrammatically represented in \Eqref{eq:tau_r_diagram_sum}. Using field-theoretic tools, the entire sum can be exactly evaluated, see \Eqref{eq:eff_tau_coupling}. The result, \Eqref{eq:Q_in_omega}, is in agreement with previous classical results. So far, we therefore have  constructed a field theory for Markovian visit probabilities that reproduces known results. In the next section we consider non-Markovian processes, and use this field-theoretic formulation to compute the perturbative corrections to the exact result \eqref{eq:Q_in_omega}. }

\section{Non-Markovian visit probabilities: A perturbative approach}
The process introduced in \eqref{eq:langevin} is driven by $\delta$-correlated noise and hence is Markovian \cite{horsthemke2006}.
As a perturbative generalisation towards non-Markovian processes, we thus extend the class of processes given by \eqref{eq:langevin} to 
\begin{align}
	\dot{x}_t = -V^{\prime}(x_t) + \xi_t + \cpl y_t \ .
	\label{eq:driven_langevin}
\end{align}
The additional driving noise $y_t$ is assumed to be stationary with zero mean and a general autocorrelation function 
\begin{equation}
C_2(t-s)=\yavg{y_s y_t}.
    \label{eq:def_C2}
\end{equation} 
By $\yavg{\bullet}$ we denote in the following averages over the path distribution of $y_t$. Importantly, $y_t$ is not required to be Gaussian such that higher non-trivial cumulants $C_n(t_2-t_1,...,t_n-t_{n-1}) = \avg{y_{t_1}...y_{t_n}}_c$ may exist. 
Such higher order cumulants enter only at perturbative order $g^n$. The correlation function $\corr{t-s}$ of the driving noise $y_t$ may  decay exponentially, such as for run-and-tumble particles in noisy environments \cite{malakar_steady_2018,dabelow_irreversibility_2019}, or algebraically.

In the following we perform a diagrammatic expansion of the visit probability $\trace(x,t)$ in the dimensionless coupling constant $\cpl$. Therefore, $\cpl$ is assumed to be small ($\cpl \ll 1$). A condition for the validity of the expansion is that cumulants of the noise $C_n(t_2-t_1,...,t_n-t_{n-1}) = \avg{y_{t_1}...y_{t_n}}_c$ are such that diagrams of the expansion are finite, for instance the diagram in Eq.~\eqref{eq:simple_loop}. In what follows, we derive the visit probability $\trace(x,t)$ averaged over both, the Gaussian white noise $\xi_t$ and the driving noise $y_t$ to first leading perturbative order $\cpl^2$. For the case of Brownian motion driven by self-correlated noise, we find expressions that depend on double integral of the cumulant $\corr{t}$, Eq.~\eqref{eq:def_upsilon}. Existence of this double integral allows for a broad class of correlation functions, even when $C_2(t)$ decays algebraically slowly at large times. Such algebraic decay occurs in processes driven by fractional Gaussian noise \cite{sevilla_generalized_2019}.

As is outlined in the App.~\ref{app:visit_driven_processes}, the visit probability conditioned on a \emph{fixed realisation} of the driving noise can still be obtained by \Eqref{eq:trace_as_avg} using the average with respect to the modified path measure
\begin{multline}
	\yavg{\cP[\phi,\tphi,\psi,\tpsi]}\\
	= \lim_{\gamma \to \infty} \int \cD[y_t] \exp\Big( -\Smob[\phi,\tphi] -\Strc[\psi,\tpsi] \\
	+ \deprate \Sdep[\varphi,\tphi,\psi,\tpsi]  
	+ \iint g y_t \tphi \partial_x \phi \Big) \cP[y_t]\, .
	\label{eq:y_cond_path_measure}
\end{multline}
As is shown in the App.~\ref{app:visit_driven_processes}, one can compute averages with respect to this $y_t$-dependent path-average, to then integrate over the path measure of $y_t$. As it turns out, one does not need to know the full path measure of $y_t$, but can instead rewrite the double average $\yavg{\avg{\bullet}}$ using the \emph{moment generating functional} of $y_t$ which is defined as
\begin{align}
       \party [g\cdot j_t] &= \int \mathcal{D} [y_t] 
       e^{ g \int \dint{t} j_t y_t } \cP[y_t] \ .
     \label{eq:partition_sum}
\end{align}
Expanding the exponential to second order, and averaging over $y_t$, gives (\cf \Eqref{eq:def_C2})
\begin{multline}
    \party [g\cdot j_t] 
    = 1 + \frac12 g^2 \iint \dint{t_1} \dint{t_2} j_{t_1} C_2(t_2 - t_1) j_{t_2} + \cO(g^3). 
    \label{eq:part_y_expansion}
\end{multline}
Therefore, we may approximate  expectation values such as the one appearing in \Eqref{eq:y_cond_path_measure} using the identity
 \begin{align}
     \yavg{\avg{\bullet}_{\cS}} = \avg{\bullet \cdot \party\left[g \iint \dint{x} \dint{t} \tphi \partial_x \phi  \right]}_{\cS}.
     \label{eq:yavg_trick}
 \end{align}
which to this perturbative order only contains $\corr{t-s}$, the correlation function of $y_t$ (\Eqref{eq:def_C2}).  

\begin{figure}[t]
\includegraphics[width=\columnwidth]{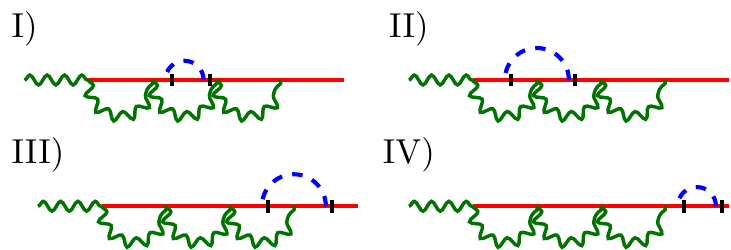}
    \caption{Diagrammatic representation of the four different kinds of terms appearing in the perturbative expansion of $\trace$ (\cf \Eqref{eq:trace_exp_in_g2} and \eqref{eq:simple_loop}). The Markovian diagrams (\cf \Eqref{eq:tau_r_diagram_sum}) are corrected by the external driving noise $\cpl \cdot y_t$, giving rise to the expansion Eq.~\eqref{eq:trace_exp_in_g2}. These four types of diagrams result in four different correction terms explicitly derived in App.~\ref{app:non_markovian_Q}, and which together give the final result for the driving noise corrected transition probability stated in \Eqref{eq:T2_formula}.}
    \label{fig:four_diagrams}
\end{figure}

Diagrammatically speaking, 
this amounts to computing the same diagrams as in \Eqref{eq:tau_r_diagram_sum}, but additionally decorating them with driving noise correlation loops represented as
\begin{align}\label{eq:simple_loop}
\cpl^2  \qquad \tikz[baseline=-2.5pt]
{
	\draw[Aactivity] (0,0) -- (2.5,0);
	\draw[Bactivity] (1.75,0) arc (0:180:.5);
	\draw[very thick] (0.85,-0.1) -- (0.85,0.1);
	\draw[very thick] (1.85,-0.1) -- (1.85,0.1);
	\node at (0.7,-.1) [anchor=north] {$y_1,s_1$};
	\node at (1.9,-.1) [anchor=north] {$y_2,s_2$};
	\node at (2.5,0) [anchor=west] {$\start$};
	\node at (0,0) [anchor=east] {$\target,t$};
	}
\end{align}
As in \Eqref{eq:vertices}, red solid lines denote bare transition probabilities. The blue dashed line connecting two internal vertices, (\cf \Eqref{eq:part_y_expansion} and App.~\ref{app:visit_driven_processes}), represents the correlation kernel $C_2(s_2,s_1)$. The two vertical bars inserted to the right of each such vertex represent the gradient operator acting on the target point of the incoming transition probability. Combinatorally, there are four ways in which these loops decorate the diagrams of the (Markovian) visit probability, which are displayed in Fig.~\ref{fig:four_diagrams}. As is shown in the App.~\ref{app:non_markovian_Q}, the new visit probability, $\trace(x,t) = \avg{\psi(x,t) \tphi(0,0) \party[ g \iint \tphi \nabla \phi ] }_{\cS}$, 
acquires a functionally similar form to the Markovian case \eqref{eq:Q_in_omega}
\begin{multline}
\trace({\start,\target,t}) \\
= \int \dbar{\omega} \frac{e^{-i\omega t}}{-i\omega}\frac{T^{(0)}(\start,\target,\omega) + \cpl^2 \transeff{\start,\target,\omega} }{R^{(0)}(\target,\omega) + \cpl^2\returneff{\target,\omega} }
+ \cO(\cpl^3)
\label{eq:nonM_trace}
\end{multline}
\new{This is a central result of the present article}.
Here, as we do from now on, we denote the (Fourier transformed) Markovian transition probability density of the undriven process in \Eqref{eq:langevin} as $T^{(0)}(x_0, x_1,\omega)$ instead of $T(x_0,x_1,\omega)$, and analogously, $R^{(0)}(x_1, \omega) = T^{(0)}(x_1,x_1,\omega)$. Further, we introduced the $g^2$ correction to the $y_t$-averaged transition probability
\begin{widetext}
\begin{multline}
	\transeff{\start,\target,\omega} =
 \iint \dint{y_1} \dint{y_2} \dbar{\tw} \left[
 \frac{ \transm{y_1,\target,\omega-\tw}}{\returnm{\target,\omega-\tw}}
 \transm{\target,y_2,\omega-\tw}- \transm{y_1,y_2,\omega-\tw}
 \right] \\
 \times\left( \partial_{y_1} \transm{\start,y_1,\omega} \right) \left(\partial_{y_2} \transm{y_2,\target,\omega} \right) \fcorr{\tw}.
 \label{eq:T2_formula}
\end{multline}
\end{widetext}
and, analogously, $\returneff{\target,\omega}  = \transeff{\target,\target,\omega} $.

In the case of a Brownian Motion additionally driven by self-correlated noise, the second order corrections given by \Eqref{eq:T2_formula} drastically simplifies as shall be demonstrated below.

\section{Example: Brownian Motion driven by self-correlated noise}
We briefly discuss the example process of active thermal Brownian motion (ATBM, \cite{Walter2021}) on a real line. In our previous article \cite{Walter2021}, we found the perturbative correction to the first-passage time moment generating function for an active thermal Ornstein Uhlenbeck process, and an active Brownian Motion on a ring. Since the distribution of first-passage times and visits are closely linked via $\trace(x,\omega) = (-i\omega)^{-1} \fptmgf(x,\omega)$, we report the visit probabilities of the latter two processes in App.~\ref{app:list_of_results}.

To illustrate the method outlined above we use the example of ATBM on a real line defined via
\begin{align}\label{eq:ATBM_Langevin}
\dot{x}_t = \xi_t + \cpl y_t
\end{align}
where $\xi_t$ is white noise of strength $\avg{\xi_s \xi_t} = 2 D_x \delta(t-s)$, and $y_t$ is a stationary, not necessarily Gaussian, driving noise with a non-white self-correlation function
$\corr{|s-t|} = \avg{y_s y_t}$. Without loss of generality, we assume $\start=0$ and $\target >0$. The transition probability $\trans{\start,\target,\omega} = \trans{x= \target-\start,\omega}$ of simple Brownian motion, $\cpl=0$, is given by
$\transm{x,t} = (4\pi D_x t)^{-1/2}\exp\left(-\frac{x^2}{4D_x t} \right)$, so that $\transm{x,\omega} = (\sqrt{-4 i \omega D_x})^{-1} \exp(-\sqrt{ -i \omega / D_x} |x| )$ and $\trace(x,t) = \new{\trace^{(0)}(x,t) =} 1 - \operatorname{erf} \left((4D_x t)^{-1/2} |x| \right)$ via \Eqref{eq:Q_in_omega} and a subsequent inverse Fourier transform.

For $\cpl \neq 0$, we compute the correction $\transeff{x,\omega}$ using Eq.~\eqref{eq:T2_formula}. As is detailed in  App.~\ref{app:T2_Brownian_Motion}, the integral simplifies drastically, and we state here only the results, which can be summarised most succinctly using a ``time-stretch function" $\Upsilon(t)$ as follows. Given the correlation function $\corr{t}$, \Eqref{eq:def_C2}, it is defined as
\begin{align}
	\Upsilon(t) = \int_{0}^{t} \dint{s} s \corr{t-s} = \int_{0}^{t} \dint{u} \int_{0}^{u} \dint{s} \corr{s},
	\label{eq:def_upsilon}
\end{align}
or, alternatively, $\partial_t^2 \Upsilon(t) = \corr{t}$ \new{with $\Upsilon(0) = 0, \partial_t \Upsilon(t) = 0$}. The second order correction to the transition probability can then be written as
\begin{multline}
	\transeff{\start,\target,\omega} = \int_{}^{} \dint{t} e^{i\omega t} \Upsilon(t) (\partial_{x_1}^2) \transm{\start,\target,t} \\
	= \int_{}^{} \dint{t} e^{i\omega t} D_x^{-1} \Upsilon(t) \,\partial_t \transm{\start,\target,t}
	\label{eq:T2_integral}
\end{multline}
where we made use of the Brownian relation $\partial_t \transm{x,t} = D_x \partial_x^2 \transm{x,t}$. 
In real time, the correction $\transeff{\start,\target,t}=D_x^{-1} \Upsilon(t) \partial_t \transm{x, t}$ can be 
absorbed into the $t$-dependence of the tree-level
as 
\begin{equation}
\trans{x,t} = \transm{x, t + \cpl^2 D_x^{-1} \Upsilon(t)} + \cO(\cpl^3).  
\end{equation}
\new{This result is in agreement with the expression for the full transition probability found in \cite{haenggi1985}, and is exact if $y_t$ further is Gaussian.}
The externally driven non-Markovian process $x_t$ has therefore  transition probabilities that are, to order $\cpl^2$, equal to those of a time-stretched Brownian motion with $ t \mapsto \tau(t) = \left( 1 + \cpl^2 (D_x t)^{-1} \Upsilon(t) \right) t$, thus $ x_t \stackrel{d}{=} \sqrt{2D_x} W_{\tau(t)}$. That, however, does not mean that the return probabilities agree between the original, externally driven non-Markovian process $x_t$ and the time-stretched Brownian motion, because of the latter not accounting for the now hidden variable $y_t$ \new{(see also the discussion in \cite{haenggi1985})}. The time-stretched Brownian motion ignores correlations between that $y_t$ and $x_t$. For example, the transition probability is no longer correctly given by the first passage and repeated return, as first passage is favoured for particular values of $y_t$ that the subsequent return does not account for.
\begin{figure*}[htp]
  \centering
  \subfigure{\includegraphics[width=.49\textwidth]{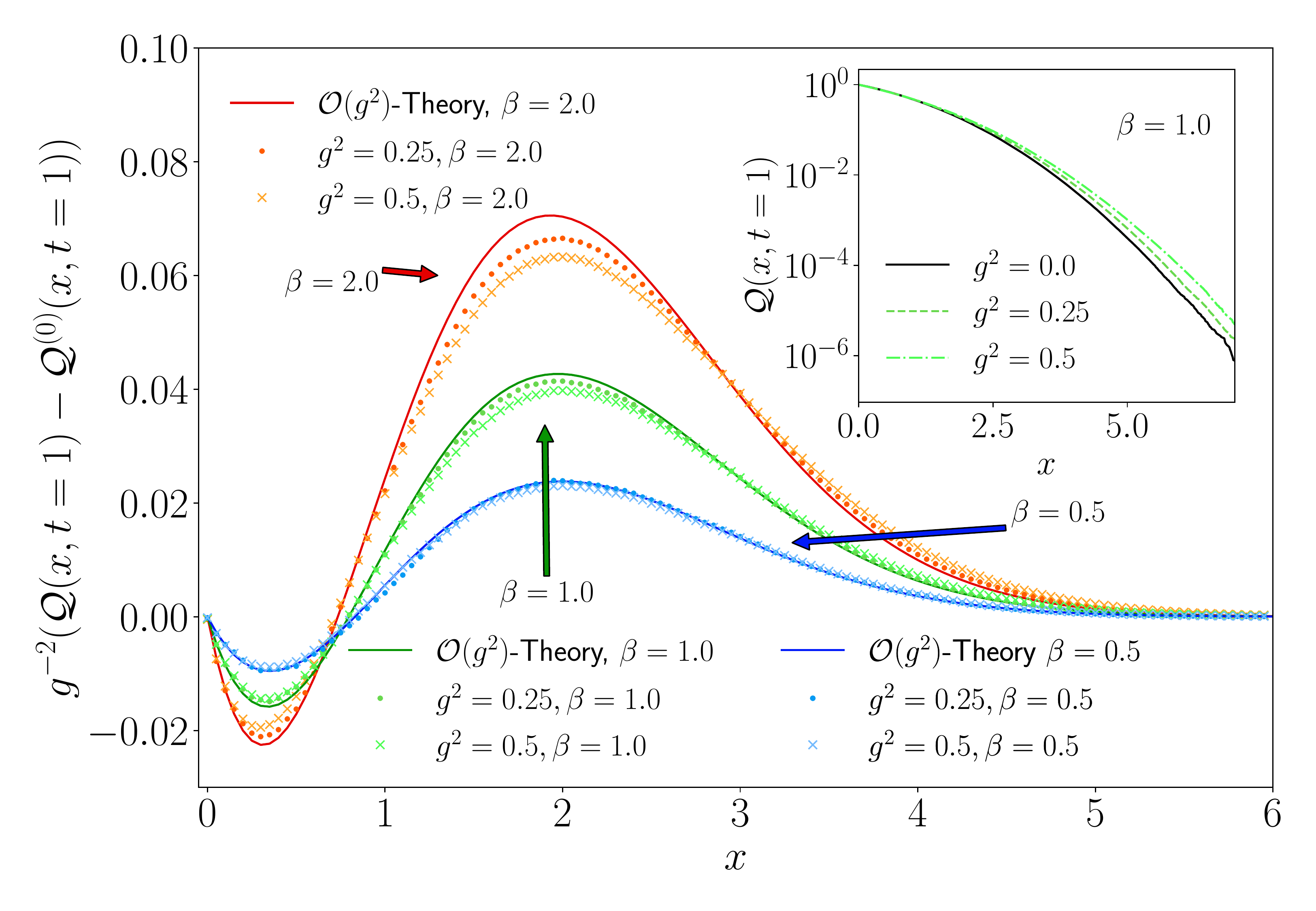}}
  \subfigure{\includegraphics[width=.49\textwidth]{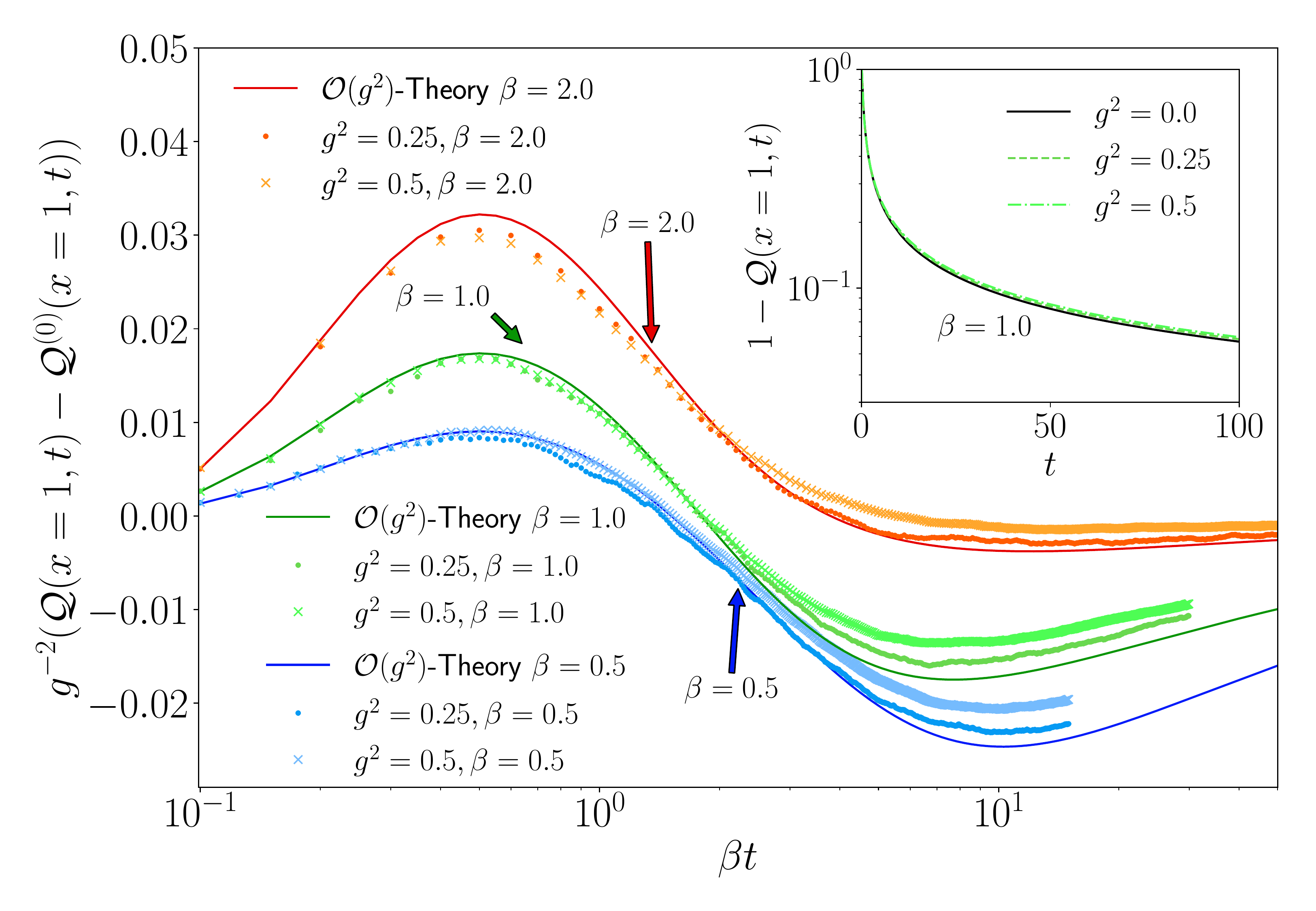}}
  \vspace{-1.7\baselineskip}
  \caption{
  Rescaled perturbative correction in $g^2$ to the visit probability $\trace(x_1, t)$ of Brownian Motion driven by correlated noise (ATBM, \cf \Eqref{eq:ATBM_Langevin}). The \textbf{left} plot shows the result for fixed time, $\trace(x_1, t=1)$, the \textbf{right} plot for fixed distance, $\trace(x_1 = 1, t)$. 
  The \textbf{inset of the left plot} shows the visit probability $\trace(x_1, t = 1)$ over distance for simple Brownian Motion ($g^2 = 0$, black line) as well as ATBM for $\beta = 1.0$ and $g^2 = 0.25$ (green dashed line), or $g^2 = 0.5$ (bright green dot-dashed line) respectively. The \textbf{inset of the right plot} shows the complement of the corresponding visit probability  $1-\trace(x_1 = 1, t)$ over time, with identical parameters and symbols. The \textbf{left (right) main panel} shows the rescaled correction to the visit probability over distance $x_1$ (rescaled time $\beta t$) for three different values of, from top to bottom, $\beta = 2.0$ (red/orange), $\beta = 1.0$ (green), and $\beta = 0.5$ (blue). Plot marks indicate the result obtained from simulation for either $g^2 = 0.25$ (circles) or $g^2 = 0.5$ (crosses). The solid lines indicate our predictions to first leading order in $g^2$ obtained by calculating the result of \Eqref{eq:nonM_trace_brownian} (\cf \Eqref{appeq:Q_line}), and numerically inverting the Fourier transform. All simulations used $D_x = D_y = 1$, and $\geq 10^6$ realisations.
  }
  \label{fig:atbm_corrections}
\end{figure*} 
Returning to \Eqref{eq:nonM_trace}, and using \Eqref{eq:T2_integral}, we slightly rephrase the result for driven Brownian Motion by explicitly expanding in $\cpl^2$ to
\begin{widetext}
\begin{multline}
    \trace(x,\omega;g)  = \trace^{(0)}(x,\omega) \left[1+ \frac{\cpl^2}{D_x}\left(\frac{\int \dint{t} e^{i\omega t} \Upsilon(t) \partial_t \transm{x,t}}{\transm{x,\omega}} -\frac{\int \dint{t} e^{i\omega t} \Upsilon(t) \partial_t \returnm{x,t}}{\returnm{x,\omega}} \right) \right] + \cO(\cpl^3)
    \label{eq:nonM_trace_brownian}
    \end{multline} 
\end{widetext}
This relation illuminates the relation between the correction to the Fourier-transformed visit probability and the correlation function of the driving noise. To obtain the visit probabilities, and derive extreme event distributions, in real time (\cf Eqs.~\eqref{eq:def_fptdist}, \eqref{eq:def_maxdist}, \eqref{eq:mean_vol}), numerical integration will be necessary in most cases.



For exponentially correlated driving noise, such as ``coloured'' or telegraphic noise, we have $\corr{t} = D_x \beta e^{-\beta |t|}$, \Eqref{eq:def_C2}, for some noise strength $D_x$ and timescale of relaxation $\beta^{-1}$. The corresponding time-stretch function is
\begin{align}
	\Upsilon(t) = \frac{D_y}{\beta} \left( e^{-\beta t} + \beta t -1 \right)
	\label{eq:Upsilon_exponential}
\end{align}
From this function alone, one can read off that for large times the driving noise effectively shifts the diffusion constant by $D_x \to D_x + \cpl^2 D_y$. At short time-scales $(t \lesssim \beta^{-1}$), however, the correction is non-trivial. 

We can further verify the validity of the expansion of the trace function $\trace(x,\omega;\cpl)$, \Eqref{eq:nonM_trace_brownian}, by numerically inverting the Fourier-transform and compare to Monte-Carlo simulations of the process \Eqref{eq:ATBM_Langevin}. To this end, we estimate numerically the probability that $x_t$ has reached $x_1$, at some given $t$ and parameters such as $\cpl$, $D_x$ and $D_y$,
and subtract from it the exact result $\trace(x,t;\cpl=0)$. Plotting this difference over $\cpl^2$ produces an estimate of the correction of $\trace(x,t;0)$ to $\trace(x,t,\cpl)$, described by \Eqref{eq:nonM_trace_brownian} to leading order (see \Eqref{appeq:Q_line} for explicit result). Fig.~\ref{fig:atbm_corrections} shows this numerically estimated correction together with the inverted correction \Eqref{eq:nonM_trace_brownian}.

The persistence exponent $\theta$ is defined as the tail exponent of the survival probability $\bP_{\text{surv}}(x,t) = 1 - \trace(x,t)  \sim t^{-\theta}$ \cite{Majumdar1999,Aurzada2015}. It is also contained in the small $\omega$ expansion of the trace as $(-i\omega)^{-1} - \trace(x,\omega) \sim \omega^{\theta - 1}$. Evaluating 
\Eqref{eq:nonM_trace_brownian} using the time-stretch function \eqref{eq:Upsilon_exponential} shows that the Markovian result of $\theta = \frac12$ \cite{Majumdar1999} does not acquire corrections \new{as is expected for short-range correlated driving noise $y_t$}.

\section{Discussion and Summary \label{sec:discussion}}
\new{In this section, we discuss our  findings in the context of the literature of (quantum) field theory and stochastic dynamics and summarise our results.}
\subsection{\new{Discussion}}
\subsubsection{Relation to Markovian techniques}

Our study is motivated by the study of complex systems comprising many interacting degrees of freedom of which we single out the slowest one as as a stochastically evolving coordinate $x_t$ (see Sec.~\ref{sec:intro}). The remaining degrees of freedom are subsumed into a bath, exerting a stochastic force onto the particle, here modelled by \Eqref{eq:driven_langevin}.

Often, the fast degrees of freedom in a complex system are assumed to evolve infinitely fast, thus rendering the stochastic evolution Markovian, since all correlations disappear within an infinitesimal time \cite{vanKampen1998}. This assumption is reflected in the mathematical structure of the usual stochastic representations, such as Langevin equations \cite{langevin_1908,gardiner_handbook_1997} or Fokker-Planck equations \cite{risken_fokker-planck_1984} which evolve locally in time, \ie with no recourse to the past evolution. In this article, we set up a field-theoretic framework for Markovian processes in Sec.~\ref{sec:ft_markov}. The ``Markovianness'' of the field-theory can be seen from the field action \eqref{eq:sdep_def} which is local in time. At this stage, the field-theory is fully equivalent to any other Markovian description, and in fact reproduces the known Markovian result for the visit probability in \Eqref{eq:Q_in_omega}.

The assumption of an infinitely fast evolving bath, however, is unphysical \cite{van_kampen_remarks_1998} and hence in each time-increment the future stochastic evolution depends on the, potentially entire, past of the trajectory. This implies that the time-local formalisms mentioned above need to be extended to include non-local time interactions (\eg to generalised Langevin equations \cite{kubo_fluctuation-dissipation_1966} or fractional diffusion equations \cite{sandev_fractional_2019}). 
In our work, this self-interaction of the process with its own past is encapsulated by the non-local contribution \eqref{eq:part_y_expansion}. The self-interaction is most clearly seen diagrammatically by the loop-diagram given in \Eqref{eq:simple_loop}.

\subsubsection{Field theory  \label{sec:discussion_qft} }
\new{
Field-theoretically inspired path integral methods are commonly used in the study of stochastic processes \cite{Chaichian2001,Feynman2010,popov_functional_2001,calvo_path_2008,meerson_path_2022}. }

\new{In this article, we begin by writing the solution of the Markovian $(\cpl = 0)$ forward equation \eqref{eq:fokker_planck}, the transition probability $T(x,t)$, as a path-integral \Eqref{eq:trans_as_avg} over fields whose distribution is given by 
 the action \Eqref{eq:smob_def}. In setting up the field theory, \Eqref{eq:joint_field_measure}, we constructed a nonequilibrium field theory using the Doi-Peliti formalism \cite{Doi1976,Peliti1985}.
In principle, the transition probability can be obtained using alternative routes, following for instance the Feynman-Kac theorem \cite{Kac1949,ito1961} which expresses the solution to parabolic PDEs such as \Eqref{eq:fokker_planck} as path integrals over  trajectories rather than fields. }
 
\new{These alternative techniques, however, do not extend to either $(i)$ the transition probability of driven, non-Markovian, processes, such as \Eqref{eq:driven_langevin}, which render the forward equation nonlocal in time, nor $(ii)$ to 
the stochastic description of the visit probability $\trace(x,t)$ which to the best of our knowledge cannot be characterised as a solution to a parabolic PDE.
This then requires two respective additional technical points with respect to alternative methods suitable to study transition probabilities of Markovian processes.}

\new{ 
In order to address the average over the driving noise $y_t$, we introduced \Eqref{eq:yavg_trick}
where we replaced the path integral over the driving noise by its moment generating function (or partition function) $\party$.
Following field-theoretic standard procedures \cite{LeBellac:1991,Ryder1996}, we expand the latter in a power series in $\cpl$. This is analogous to the perturbative treatment of self-interaction in classical field theories.
}

\new{Secondly, in order to cast the visit probability $\trace(x,t)$ (which does not readily follow from a Fokker-Planck equation) into a field theory, we used the tracing mechanism and its Doi-Peliti formulation to track the volume explored via the auxiliary fields $\psi, \tpsi$. This formalism (introduced in \cite{nekovar_field-theoretic_2016}) does not correspond to a classical field theoretic technique, but is rooted in the study of reaction-diffusion methods with field-theoretic methods.
}

\subsubsection{Stochastic Dynamics}

\new{
The problem of finding the transition probability, let alone the extreme event distribution, of non-Markovian processes has been a long-standing problem in stochastic dynamics.
In terms of computing the transition probability, our field-theoretic approach recovers results known in the literature \cite{haenggi1985} in which non-$\delta$-correlated driving noises are also treated using functional methods (see also \cite{jung_dynamical_1987,hanggi_reaction-rate_1990} for a similar discussion of first-passage times). 
 However, for non-Markovian processes the \emph{visit probability} does not straight-forwardly follow from the transition probability. 
 The visit probability $\trace(x,t)$ therefore represents the central result of the present work.
}

\subsection{\new{Summary}}
In the present work we have established a method to compute the visit probability (the complement of the survival probability) for random motion in a one-dimensional potential, as defined in \Eqref{eq:driven_langevin}.

By mapping the problem to a field-theory we systematically compute corrections to the Markovian result \eqref{eq:Q_in_omega} to any order in the coupling $\cpl$. The leading order correction is of the form $\eqref{eq:nonM_trace}$, which involves the $\cpl^2$-corrected transition probability given in \eqref{eq:T2_formula}. Generally, to compute contribution to order $g^n$, it is sufficient to know the Markovian transition probability and the $n$-point function of the driving noise $y_t$. In the absence of an external potential, the expressions reduce to \eqref{eq:nonM_trace_brownian} which depends on the twice-integrated driving noise correlation $\Upsilon(t)$, \Eqref{eq:def_upsilon}.

By casting the problem in a field-theoretic language, we replaced the single degree of freedom by a field $\phi$ representing the full density. This allows us to extend the model to the case of many, potentially interacting, random walkers. Also, the Doi-Peliti framework allows for the inclusion of potentials. To the best of our knowledge, this is the first time that both Doi-Peliti and Martin-Siggia-Rose have been used simultaneously to construct an action.

Overall, we have established a method which further lays bare the interplay between non-Markovianness and extreme events in stochastic processes.

\section*{Acknowledgements}
BW and GP would like to thank the Francis Crick Institute for their hospitality. BW would like to thank Andrea Gambassi for insightful discussions and comments.
BW acknowledges financial support from MIUR-PRIN project  ``Coarse-grained  description  for  non-equilibrium systems and transport phenomena (CO-NEST)'' n.  201798CZL.
GS was supported by the Francis Crick Institute which receives its core funding from Cancer Research UK (FC001317), the UK Medical Research Council (FC001317), and the Wellcome Trust (FC001317).

\normalem
%


\appendix 
\begin{widetext}

\section{Tracing mechanism}
\label{app:tracing_mechanism}
The visit probability for Markovian processes (\cf \eqref{eq:langevin}), given in \eqref{eq:trace_as_avg}, is a result that can be derived in various ways (indirectly, via the first-passage time distribution, $\partial_t \trace (x,t)$, this result has been found in, \eg, \cite{darling_first_1953} using a renewal type approach). Here, we derived the result by taking the continuum limit of a discrete reaction-diffusion process designed to track visits, which we refer to as ``tracing mechanism'' and which has been introduced in \cite{nekovar_field-theoretic_2016, bordeu_volume_2019}.

To describe the tracing mechanism, we consider a coarse-grained version $\tx_t$ of the stochastic process $x_t$ of \Eqref{eq:langevin}, which takes values only on a lattice $\latt \mathbb{Z}$ where $\latt$ is the lattice-spacing, so that formally $\tx_t = \latt [ \latt^{-1} x_t]$ where $[x]$ rounds to the nearest integer. At any time, the random walker attempts to deposit a \emph{trace} at $\tx_t$ with Poissonian rate proportional to $\gamma$. Each lattice site, however, has a ``carrying capacity'' $\overline{n}_0$ which limits the number of trace particles that can be deposited at this site. If at a site $\overline{n}_0$ trace particles have already  been deposited, any further deposition is suppressed. Hence, the number of trace particles deposited at any lattice site is an integer bound above by $\overline{n}_0$. Taking $\gamma \to \infty$, the particle  deterministically deposits trace particles at any site visited for the first time. In the limit of $\latt \to 0$, the process $\tx_t$ tends to $x_t$, and the expected number of trace particles at a site, divided by $\overline{n}_0$, converges to $\trace(x,t)$.

Cast into the language of reaction diffusion processes, we have at each lattice site $i$
\begin{align}
    W_{\new{i}} + n T_{\new{i}} &\stackrel{\gamma}{\longrightarrow} W_{\new{i}} + (n+1) T_{\new{i}} & n < \new{\overline{n}_0} \label{appeq:chem_1}\\
    T_{\new{i}} &\stackrel{\eps}{\longrightarrow} \varnothing  \label{appeq:chem_2}
\end{align}
where particles of species $W$ (``walkers'') deposit particles of species $T$ (``traces'') at rate $\gamma$, provided their number does not surpass the carrying capacity $\overline{n}_0$.  Meanwhile $W$ diffuses according to (a discretised form of) \eqref{eq:langevin}, $T$ remains at a given site and ``evaporates'' with rate $\eps$, later sent to zero.

The Doi-Peliti formalism \cite{Doi1976,Peliti1985} describes the continuum limit of particle densities in a reaction-diffusion system, such as defined in Eqs.~\eqref{appeq:chem_1}, \eqref{appeq:chem_2}, by mapping the problem onto a non-equilibrium field theory, where each particle species corresponds to a pair of fields \new{(see \cite{nekovar_field-theoretic_2016} for details)}. The \new{local density of} walkers \new{$\latt^{-1}W_i$} is mapped to $\phi(x,t), \tphi(x,t)$ (referred to as annihilation and creation fields, respectively), and the  trace particle \new{density $\lim_{\latt \to 0} \latt^{-1} T_i $} to $\psi(x,t), \tpsi(x,t)$. As shown in \cite{nekovar_field-theoretic_2016,bordeu_volume_2019}, the joint distribution of the four fields then follows from the Doi Peliti framework to be distributed according to 
 the action given in Eqs.~\eqref{eq:smob_def}, \eqref{eq:joint_field_measure}, and \eqref{eq:sdep_def}.
 \new{In order to turn the density of tracers into a probability of visit, it needs to be divided by the normalising density $\carryingCap $ corresponding to the continuum limit of $\overline{n}_0$. Hence, the field-theoretic formula for the visit probability, \Eqref{eq:trace_as_avg}, contains a prefactor of $\carryingCap^{-1}$.}

\section{Field-theoretic Calculation of Markovian visit probability}
\label{app:markovian_visit_p}
We derive \Eqref{eq:Q_in_omega}, the expression for the visit probability in the Markovian case. 
First, we consider the field theory in  the case of $\gamma = 0$, when the probability measure \eqref{eq:joint_field_measure} is Gaussian.

We introduce the forward and backward operators $\LL, \LL^{\adj}$, 
\begin{align}
	\label{eq:forward}
	\LL(x) &= V^{\prime\prime}(x) + V^{\prime}(x)\partial_x + D_x \del_x^2\ ,\\
		\LL\adj(x) &=  - V^{\prime}(x)\partial_x + D_x \del_x^2
\end{align}
associated to \eqref{eq:langevin} and generating the corresponding Fokker Planck Equation 
\begin{align}
		\nonumber
		\partial_t \trans{x,t} &= \LL \trans{x,t} \\
		T(x,t=t_0) &= \delta(x-x_0)
		\label{eq:forward_equation}
\end{align}
with $\trans{x,t}$ the transition probability of the process.
The forward and backward operator have
a set of eigenfunctions 
\begin{align}
	\LL u_n(x) &= -\lambda_n u_n(x) \\
	\LL\adj v_n(x) &= -\lambda_n v_n(x)
	\label{}
\end{align}
with eigenvalues $0 \leq \lambda_0 < \lambda_1 ...$.
The eigenfunctions are $L^2$-normalised to satisfy the orthonormal relation
\begin{align}
	\int \dd{x} u_m(x) v_n(x) = \delta_{mn}, \label{eq:delta_uv_1}
\end{align}
and since they form a complete set in $L^2$ they further satisfy \cite{risken_fokker-planck_1984}
	\begin{align}
	\sum_n v_n(x_1) u_n(x_2) = \delta(x_1-x_2) \label{eq:delta_uv_2}. 
	\end{align}
Thus, every field $\phi, \tphi, \psi,\tpsi$ has a unique decomposition into the $u_n(x), v_n(x)$ in space. Together with a Fourier transform in time, we then write
\begin{align}
    	\varphi(x,t) &= \int_{}^{} \dbar \omega \sum_{k}^{} \varphi_k(\omega)  u_k(x) e^{-i \omega t}  \label{varphi_trafo}\\
	\tilde{\varphi}(x,t) & = \int \dbar \omega\sum_{k}^{}  \tilde{\varphi}_k(\omega) v_k(x) e^{-i \omega t}
	\label{varphi_tilde_trafo},
\end{align}
and analogously for $\psi,\tpsi$ with coefficients $\psi_k(\omega), \tpsi_k(\omega')$, respectively. This \emph{mode transform} diagonalises the non-perturbative parts of the action, $\Smob$ and $\Strc$ (\cf \Eqref{eq:joint_field_measure}), which read
\begin{align}
	\Smob[\phi,\tphi] &= \int \dbar{\omega} \sum_n \tphi_n(-\omega)\left(-i\omega + \lambda_n \right)\phi_n(\omega) \label{eq:smob_diag} 
	\\
	\Strc[\psi,\tpsi] &= \int \dbar{\omega} \sum_n \tpsi_n(-\omega)\left( -i\omega + \eps \right)\psi_n(\omega)
	\label{eq:strc_diag}.
\end{align}
For $\gamma = 0$, the measure in \Eqref{eq:joint_field_measure} is Gaussian and the (Fourier transformed) bare propagators of both fields, $\avg{\phi \tphi}$ and $\avg{\psi \tpsi}$, therefore immediately follow from Eqs.~\eqref{eq:smob_diag} and \eqref{eq:strc_diag} using standard path integral techniques \cite{Taeuber:2014},
\begin{align}
   	\langle \varphi_n(\omega') \tphi_m(\omega) \rangle = \frac{\delta_{m,n} \delta(\omega + \omega')}{-i \omega' + \lambda_n} = 
	\tikz[baseline=-2.5pt]{\draw[Aactivity] (0,0) -- (1.5,0) node [at start, above] {$(m,\omega')$} node [at end, above] {$(n,\omega)$};} \label{eq:bare_phi_corr}\\
   	\langle \psi_n(\omega') \tpsi_m(\omega) \rangle = \frac{\delta_{m,n} \delta(\omega + \omega')}{-i \omega' + \eps} = 
	\tikz[baseline=-2.5pt]{\draw[Bsubstrate] (0,0) -- (1.5,0) node [at start, above] {$(m,\omega')$} node [at end, above] {$(n,\omega)$};}, \label{eq:bare_psi_corr}
\end{align}
where we introduced a diagrammatic representation for both bare propagators. These propagators are commonly interpreted as the (linear) response functions \cite{Taeuber:2014}.
Using \Eqref{eq:trans_as_avg}, and transforming back into real space and time using Eqs.~\eqref{varphi_trafo} and \eqref{varphi_tilde_trafo}, we obtain the transition probability
\begin{align}
	\trans{x,t} = \sum_n v_n(x_0)u_n(x) e^{-\lambda_n(t-t_0)} \Theta(t-t_0)
	\label{eq:trans_bare_explicit}
\end{align}
where $\Theta(t)$ is the Heaviside $\Theta$-function. Crucially, we made use of the property \cite{cardy1999,Taeuber:2014}
\begin{align}
	\avg{\phi}_{\cS}= 0
	\label{}
\end{align}
such that $\avg{\phi(1+\tphi)} = \avg{\phi \tphi}$.

We next consider the expectation \eqref{eq:trans_as_avg} in the case of $\gamma \neq 0$ when the non-linear contributions of $\Sdep$, \Eqref{eq:sdep_def}, enter. Each of the four vertices is diagrammatically represented as
\begin{align}
    \tikz[baseline=-2.5pt]{
    \node at (0,.3) {$\tau$};
    \draw[Aactivity] (0,0) -- (.3,0) ;
    \draw[Bsubstrate] (-.3,0) -- (0,0); 
    }\qquad
     \tikz[baseline=-2.5pt]{
      \node at (0,.3) {$\sigma$};
    \draw[Aactivity] (-.3,0) -- (.3,0) ;
    \draw[Bsubstrate] (0,0) -- (-.3,-.2);
    }\qquad
    \tikz[baseline=-2.5pt]{
     \node at (0,.3) {$-\lambda$};
    \draw[Aactivity] (0,0) -- (.3,0) ;
    \draw[Bsubstrate] (-.3,0) -- (0,0);
    \draw[Bsubstrate] (0,0) -- (.3,-.2);
    } \qquad 
      \tikz[baseline=-2.5pt]{
       \node at (0,.3) {$-\kappa$};
    \draw[Aactivity] (-.3,0) -- (.3,0) ;
    \draw[Bsubstrate] (-.3,-.2) -- (0,0);
    \draw[Bsubstrate] (0,0) -- (.3,-.2);}
\end{align}
and enters into the action multiplied by $\gamma$. \new{Following Refs.~\cite{nekovar_field-theoretic_2016,bordeu_volume_2019}, we introduce the \emph{carrying capacity density} $\carryingCap$ which is the continuum limit of $\latt^{-1} \overline{n}_0$, where $\overline{n}_0$ is the maximal number of  trace particles which can simultaneously be deposited at a single site. At bare level, the carrying capacity enters in the couplings via the relation 
\begin{align}
\lambda = \kappa = \carryingCap^{-1} \tau = \carryingCap^{-1} \sigma \ .
\label{appeq:bare_values}
\end{align}}
It follows that the diagrammatic expansion of  the trace function \eqref{eq:trace_as_avg}, which counts the average number of tracer particles at a site, needs to be normalised with $\carryingCap^{-1}$ in order to indicate the \emph{visit probability}, and hence
\begin{align}
	\trace(x,t) &=   \carryingCap^{-1} \lim_{\gamma \to \infty}
    \tikz[baseline=-2.5pt]{
    \draw[Aactivity] (0,0) -- (.7,0) node [at end, above] {$(x_0, t_0)$}; 
    \draw[Bsubstrate] (-.7,0) -- (.,0) node [at start, above] {$(x,t)$};
    \fill (0,0) circle(3pt);} \\
    \int \dint{t} e^{i\omega t} \trace(x,t) &= \carryingCap^{-1} \avg{\psi(x)\tpsi(x)}\left( \lim_{\gamma \to \infty} \tau_R(\omega)\right) \avg{\phi(x,\omega) \tphi(\start,\omega)} \\
    &= \carryingCap^{-1} \frac{1}{-i\omega + \eps}\left( \lim_{\gamma \to \infty} \tau_R(\omega)\right) T(x_0, x, \omega)
\end{align}
where the central dot in the diagram stands for the renormalised coupling $ \tau_R$, and $T(x_0,x,\omega)$ is the Fourier transform of the transition probability. This renormalisation of $\tau$ is given by the diagrammatic expansion of the amputated vertex
\begin{align}
     \tau_R 
     = \tikz[baseline=-2.5pt]{
    \draw[Aactivity] (0,0) -- (.2,0) ; 
    \draw[Bsubstrate] (-.2,0) -- (.,0) ;
    \fill (0,0) circle(3pt);}
    = \gamma \tikz[baseline=-2.5pt]{
    \node at (0,.3) {$\tau$};
    \draw[Aactivity] (0,0) -- (.2,0) ;
    \draw[Bsubstrate] (-.2,0) -- (0,0); 
    }
    + \gamma^2
    \tikz[baseline=-2.5pt]{
    \node at (0,.3) {$-\lambda$};
    \node at (0.5,.3) {$\sigma$};
    \draw[Bsubstrate] (-0.3,0) -- (-0,0) ;
    \draw[Aactivity] (0,0) -- (.65,0) ;
    \draw[Bsubstrate] (0.0,0) arc (180:360:.25);
    }
     + \gamma^3
    \tikz[baseline=-2.5pt]{
        \node at (0,.3) {$-\lambda$};
    \node at (0.5,.3) {$-\kappa$};
    \node at (1.0,.3) {$\sigma$};
    \draw[Bsubstrate] (-0.3,0) -- (-0,0) ;
    \draw[Aactivity] (0,0) -- (1.15,0) ;
    \draw[Bsubstrate] (0.0,0) arc (180:360:.25);
    \draw[Bsubstrate] (0.5,0) arc (180:360:.25);
    }+ ...
    \label{appeq:tau_r_diagram_sum} 
    \end{align}
The only diagrams contributing to this expansions are chains of the loop-diagram $\tikz[baseline=-5pt]{
    \draw[Aactivity] (0,0) -- (.5,0) ;
    \draw[Bsubstrate] (0.0,0) arc (180:360:.25);}$, refered to in the following as a "bubble".
Considering the expansion in Fourier and mode transform, Eqs.~\eqref{varphi_trafo} and \eqref{varphi_tilde_trafo} (App.~\ref{app:tau_renorm} 
for details), each diagram factorises into a product over the bubbles and can hence be evaluated using a geometric or Dyson sum  (App.~\ref{app:tau_renorm}
 for derivation) resulting in
    \begin{align}
	    \tau_R(\omega_1) 
	    = \frac{\gamma\tau}{1+\gamma\kappa\return{x_1,\omega_1}}
	    \label{appeq:eff_tau_coupling}
    \end{align}
    such that the effective trace function, Fourier transformed, is
    \begin{align}\label{appeq:Q_in_omega}
	    \int \dint{t}e^{i\omega t} \trace(x,t)  
	    =  \lim_{\gamma \to  \infty} \frac{1}{-i\omega + \eps} \frac{\gamma \tau \carryingCap^{-1}}{1+\gamma \kappa \return{x_1,\omega+i\eps}}\trans{\start,\target,\omega}  
	    = \frac{1}{-i\omega + \eps} \frac{\trans{\start,\target,\omega} }{ \return{x_1,\omega+i\eps}}    \end{align}
    where we made use of time-translational invariance to write the Fourier-transform in one frequency only. 
    The couplings $\tau/\kappa$ in \Eqref{appeq:eff_tau_coupling} cancel with $n_0^{-1}$  following their bare values, \Eqref{appeq:bare_values}.
    This then leads to the central Markovian result for the trace function
    \begin{align}
	    \trace(x,t) &= \int \dbar{\omega} e^{-i\omega t}\frac{\trans{\start,\target,\omega}}{(-i\omega)\return{\target, \omega}}     
    \end{align}
    where we have tacitly taken the limit $\eps \to 0$.

\section{Visit probability for driven process}
\label{app:visit_driven_processes}
In this appendix, we provide some technical details to the derivation of the key result \ref{eq:nonM_trace} and $\ref{eq:T2_formula}$ which together provide the visit probability for non-Markovian processes of the form \eqref{eq:driven_langevin}.

There are three technical steps: First, we consider the perturbative correction of the visit probability in the presence of a fixed, but random, realisation of the driving noise $y_t$. Secondly, we average over all such realisations of $y_t$. This then leads to a large set of correction terms which we interpret diagrammatically and which, thirdly, we evaluate to leading perturbative order.

\subsection{Averaging general observables over driving noise}
\begin{figure}
\includegraphics[width=.4\textwidth]{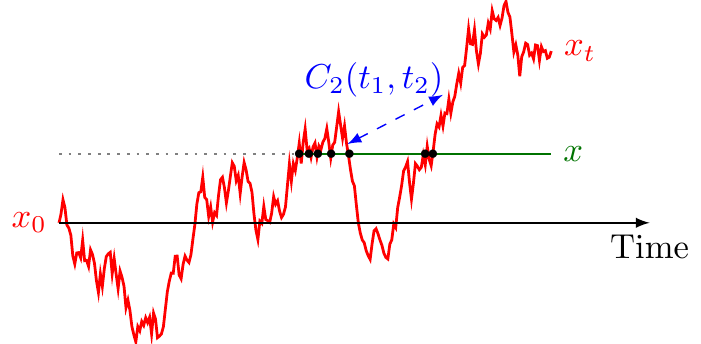}
    \caption{The random walker $x_t$ (red solid path) travels from $x_0<x_1$ to $x_t>x_1$, thereby passing $x_1$ (green solid line) infinitely often (black dots). If the random walker is additionally driven by self-correlated noise (\cf \Eqref{eq:driven_langevin}), this induces correlations between increments at any different times $t_1,t_2$ (blue dashed line). Meanwhile the perturbative expansion in $\gamma$ (\cf \Eqref{eq:tau_r_diagram_sum}) tracks the probability of all possible transitions $\trans{x_1,t}$ and subsequent returns from $x_0$ to $x_1$, the second perturbative expansion in $\cpl$ includes the effect of correlated increments. If $\cpl=0$, the Markovian case, the process undergoes renewal at every return at $x_1$, thus rendering the results such as \Eqref{appeq:Q_in_omega} \emph{exact}.}
    \label{fig:plot_random_walker}
\end{figure}
The presence of the autocorrelated driving noise $y_t$ affects the transition and return probabilities of the random walker $x_t$ (see Fig.~\ref{fig:plot_random_walker}). Conditioned on a fixed realistion of $y_t$, the forward operator (\cf Eqs.~\eqref{eq:langevin}, \eqref{eq:forward_equation}, \eqref{eq:forward} and \eqref{eq:driven_langevin}) is shifted by
\begin{align}
	\LL \mapsto \LL + \cpl y_t \del_x \varphi(x,t)
	\label{}
\end{align}
Hence, the $y_t$-conditioned random walker action \Eqref{eq:smob_def}, which is essentially of the form $\tphi(\partial_t - \LL)\phi$, is shifted also,
\begin{align}
	\Smob[y_t] = \iint \dd{x} \dd{t} \tphi ( \del_t - \LL -\cpl y_t \del_x)\phi.
	\label{eq:y_dep_action}
\end{align}
To obtain the $y$-averaged joint distribution of the four fields, one needs to evaluate (\Eqrefs{eq:joint_field_measure}, \eqref{eq:y_cond_path_measure})
\begin{align}
	\yavg{\cP[\phi,\tphi,\psi,\tpsi]}
	= \lim_{\gamma \to \infty} \int \cD[y_t] \exp\Big( -\Smob[\phi,\tphi] -\Strc[\psi,\tpsi] + \deprate \Sdep[\varphi,\tphi,\psi,\tpsi]  
	+ \iint g y_t \tphi \partial_x \phi \Big) \cP[y_t]\,,
	\label{appeq:y_cond_path_measure}
\end{align}
where $g y_t \tphi \partial_x \phi$ may be interpreted as a new vertex, (see \Eqref{eq:simple_loop}. Whilst we were able to compute the result exactly in $\gamma$, we need to resort to perturbative methods to approximate in small couplings $\cpl$.
Although the path-integral in \eqref{eq:y_cond_path_measure}  appears to require complete knowledge of the full path measure $\cP[y_t]$ of the driving noise $y_t$, we can relax this requirement by introducing the normalised partition function (or moment generating function) of $y_t$,
\begin{align}
    \party [g\cdot j_t] &= \int \cD[y_t] \exp\left( g\int \dint{t} j_t y_t \right) \cP[y_t] \\
    &= 1 + \frac12 g^2 \iint \dint{t_1} \dint{t_2} j_{t_1} C(t_2 - t_1) j_{t_2} + \cO(g^3)
\end{align}
where we ignore terms of higher perturbative order. We may thus replace the $y_t$-integration in \Eqref{eq:y_cond_path_measure} by  inserting $\party\left[g \iint \dint{x} \dint{t} \tphi \partial_x \phi  \right]$ into expectations over the fields and consequently evaluate  double averages via
\begin{align}
    \yavg{\avg{\bullet}}_{\cS+ gy} = \avg{\bullet \cdot \party\left[g \iint \dint{x} \dint{t} \tphi \partial_x \phi  \right]}_{\cS}.
    \label{appeq:yavg_trick}
\end{align}
For example, the transition probability, averaged over all driving noises $y_t$, acquires a  correction term which up to $g^2$ reads
\begin{align}\label{eq:trans_to_g2}
    \trans{\target,t} &=      \Big\langle \phi(\target,t)  \tphi(\start,0) \Big \rangle \\
    &+ g^2 \iint \dint{y_1}\dint{s_1} \dint{y_2} \dint{s_2} C_2(s_2-s_1)
    \Big\langle \phi(\target,t) 
    \tphi(y_1,s_1) 
    \partial_{y_1} 
    \phi(y_1,s_1) \tphi(y_2,s_2) \partial_{y_2} \phi(y_2,s_2) \tphi(\start,0) \Big \rangle_{\cS} 
    + \cO(g^4) \nonumber
\end{align}
The remaining field average in \eqref{eq:trans_to_g2} can now be evaluated normally, using
standard Wick product rules (see App.~\ref{app:markovian_visit_p}), \Eqref{eq:trans_as_avg}, and the non-driven path measure \eqref{eq:joint_field_measure}. 
A straightforward calculation shows that the only non-vanishing 
correction to order $g^2$ in \Eqref{eq:trans_to_g2} 
is
\begin{align}
    g^2 \int_0^t \dint{s_1} \int_0^{s_1} \dint{s_2} \iint \dint{y_1} \dint{y_2} 
    \trans{y_1,\target,t-s_1} 
    \left( \partial_{y_1} \trans{y_2,y_1,s_1-s_2}\right) C_2(s_1-s_2) \left(\partial_{y_2} \trans{x,y_2,s_1}\right)
    \label{eq:g2_corr_trans_wick}\,,
\end{align}
using 
$\langle \phi(x_1,t)\tphi(y_1,s_1)\rangle = \trans{y_1,x_1,t-s_1}$,
$\langle \partial_{y_1} \phi(y_1,t)\tphi(y_2,s_1)\rangle = \partial_{y_1} \trans{y_2,y_1,s_1-s_2}$ 
\emph{etc}.
Here, and in particular for more cumbersome expressions, it is advantageous to use diagrammatics to keep track of perturbative correction terms. The correction of the transition probability in \Eqref{eq:g2_corr_trans_wick} is represented as
\begin{align}\label{appeq:simple_loop}
\cpl^2  \qquad \tikz[baseline=-2.5pt]
{
	\draw[Aactivity] (0,0) -- (2.5,0);
	\draw[Bactivity] (1.75,0) arc (0:180:.5);
	\draw[very thick] (0.85,-0.1) -- (0.85,0.1);
	\draw[very thick] (1.85,-0.1) -- (1.85,0.1);
	\node at (0.7,-.1) [anchor=north] {$y_1,s_1$};
	\node at (1.9,-.1) [anchor=north] {$y_2,s_2$};
	\node at (2.5,0) [anchor=west] {$\start$};
	\node at (0,0) [anchor=east] {$\target,t$};
	}
\end{align}
As in \Eqref{eq:bare_phi_corr}, red solid lines denote bare transition probabilities. The blue dashed line connecting two internal vertices, \Eqref{eq:y_cond_path_measure}, represents the correlation kernel $C_2(s_2,s_1)$. The two vertical bars inserted to the right of each such vertex represent the gradient operator acting on the target point of the incoming transition probability.  As usual for Feynman diagrams, external 
fields depend on fixed parameters ($\start,\target,t$), whilst internal 
fields depend on variables to be integrated over (\eg $y_1,s_1,y_2,s_2$). In addition to providing a better overview of terms arising in the perturbative expansion, diagrams also act as graphical cues illustrating  how the driving noise induces memory into the evolution of $x_t$.

\subsection{The driving noise averaged visit probability}
Following \Eqref{eq:yavg_trick}, we are able to evaluate driving noise averaged observables. In order to derive the central results of Eqs.~\eqref{eq:nonM_trace} and \eqref{eq:T2_formula}, it remains to evaluate the driving noise averaged visit probability
\begin{align}
    \trace(x,t) &= \avg{\psi(x,t) \tphi(0,0) \party[g \int \tphi \nabla \phi]}_{\cS}\\
    &= \Bigg\langle \psi(\target, t) \phi(\start,0)
    \Bigg(1 + \cpl^2 \iint \dint{y_{1,2}} \dint{s_{1,2}} \tphi_1 \partial_{y_1} \phi_1 
    C_2(s_1-s_2)
    \tphi_2 \partial_{y_2} \phi_2  \Bigg) \Bigg\rangle_{\cS}
    + \cO(\cpl^3)
    \label{eq:trace_exp_in_g2} 
\end{align}
where we use $\phi_i = \phi(y_i,s_i)$ for brevity. Each term appearing in the $\gamma$-perturbative expansion of the trace function (\cf \eqref{eq:tau_r_diagram_sum}) is additionally corrected to order $\cpl^2$ by replacing two internal $\phi \tphi$ propagators by two ``$y_t$-driven propagators'' $\tphi\partial_x \phi$ and connecting them with the two-point correlator $C_2$ of $y_t$. As can be seen easiest diagrammatically, all possible corrections fall into one of the four categories shown in Fig.~\ref{fig:four_diagrams}. They are classified according to whether both $\cpl$-vertices couple to the same or different propagator of a transition or a return. It is simplest to compute the four contributions in frequency rather than direct time as the loops factorise. The calculation itself is given in App.~\ref{app:non_markovian_Q}.

\section{Renormalisation of transmutation rate\label{app:tau_renorm}}

In this appendix, we derive the result for the renormalisation of the coupling $\tau$ which is stated in \Eqref{eq:eff_tau_coupling}. To simplify the computation, we perform the calculation in $x,\omega$ variables, \ie in real space and Fourier transformed time. The renormalisation is given by (\cf \Eqref{eq:tau_r_diagram_sum})
\begin{align}
\label{tau_R_diagrams}
    \tikz[baseline=-2.5pt]{
    \draw[Aactivity] (0,0) -- (.7,0) ; 
    \draw[Bsubstrate] (-.7,0) -- (.,0) ;
    \fill (0,0) circle(3pt);}
    &= \gamma \tikz[baseline=-2.5pt]{
    \node at (0,.3) {$\tau$};
    \draw[Aactivity] (0,0) -- (.5,0) ;
    \draw[Bsubstrate] (-.5,0) -- (0,0); 
    }
    + \gamma^2
    \tikz[baseline=-2.5pt]{
    \node at (0,.3) {$-\lambda$};
    \node at (0.5,.3) {$\sigma$};
    \draw[Bsubstrate] (-0.3,0) -- (-0,0) ;
    \draw[Aactivity] (0,0) -- (.8,0) ;
    \draw[Bsubstrate] (0.0,0) arc (180:360:.25);
    }
     + \gamma^3
    \tikz[baseline=-2.5pt]{
        \node at (0,.3) {$-\lambda$};
    \node at (0.5,.3) {$-\kappa$};
    \node at (1.0,.3) {$\sigma$};
    \draw[Bsubstrate] (-0.3,0) -- (-0,0) ;
    \draw[Aactivity] (0,0) -- (1.3,0) ;
    \draw[Bsubstrate] (0.0,0) arc (180:360:.25);
    \draw[Bsubstrate] (0.5,0) arc (180:360:.25);
    }+ ...
    \end{align}
    which is a diagrammatic representation of the terms arising in the path-integrated average of the visit probability, \Eqref{eq:trace_as_avg}, when expanding in $\deprate$, 
\begin{multline}
	\carryingCap^{-1} \avg{\psi(\target,\omega_1)\tphi(\start,\omega_0)}_{\cS} =   \carryingCap^{-1} \gamma \tau \iint \dd{z} \dbar{\omega'} \avg{\psi(\target,\omega_1)\tpsi(z,\omega')\phi(z,\omega')\tphi(\start,\omega_0)}_{\cS;\deprate=0}\\
	- \carryingCap^{-1} \gamma^2 \lambda \sigma \iint \dd{z_1}\dd{z_2}\dbar{\omega'_1}\dbar{\omega'_2}\dbar{\omega_1{''}}\dbar{\omega_2{''}} \\
	\times\avg{\psi(\target,\omega_1) \tpsi(z_1,-\omega'_1-\omega''_1)\psi(z_1,\omega_1{''})\phi(z_1,\omega'_1)\tpsi(z_2,\omega_2{''}) \tphi(z_2,\omega'_2)\phi(z_2,-\omega'_2-\omega''_2)\tphi(\start,\omega_0)  }_{\cS;\deprate=0} + \cdots.
	\label{appeq:path_avg_1}
\end{multline}
Crucially, the averages $\avg{\bullet}_{\cS;\deprate=0}$ are taken over Gaussian random variables since for $\deprate =0$ the path action  (\Eqref{eq:joint_field_measure}) is bilinear. Thus, Wick's theorem (\eg \cite{LeBellac:1991}) applies and all averages in \Eqref{appeq:path_avg_1} decompose into products of two-point functions. The only such Gaussian two-point functions  which do not vanish are
\begin{align}
	\label{app:eq:wick_sum}
	\avg{\phi(z_1,\omega_1)\tphi(z_0,\omega_0)}_{\cS;\deprate=0} &= \trans{z_0,z_1;\omega_1}\deltabar(\omega_0 + \omega_1) = \int \dint{t}  e^{i\omega_1 (t -t_0)}\trans{z_0,z_1;t} \deltabar(\omega_0 + \omega_1)\\
	\label{appeq:Psipsi_propagator}
	\avg{\psi(z_1,\omega_1)\tpsi(z_0,\omega_0)}_{\cS;\deprate=0}&= \frac{\delta(z_1-z_0)\deltabar(\omega_0+\omega_1)}{-i\omega_1 + \eps} = \iint \dint{t_0} \dint{t} e^{i\omega_1 t +i \omega_0t_0} \Theta(t-t_0)\delta(z_1-z_0) e^{-\eps(t-t_0)}
\end{align}
The second correlator intuitively characterises the behaviour of the trace which, once deposited at $z_0$ at time $t_0$, remains there for an infinitely long time, as $\eps \to 0$. Equipped with these correlators, the non-vanishing contributions to the averages appearing in \Eqref{appeq:path_avg_1} are, following Wick's theorem,
\begin{align}
	\carryingCap^{-1} \gamma \tau \iint \dd{z} \dbar{\omega'} \avg{\psi(\target,\omega_1)\tpsi(z,\omega')\phi(z,\omega')\tphi(\start,\omega_0)}_{\cS;\deprate=0} &= \carryingCap^{-1}  \gamma \tau \iint \dd{z} \dbar{\omega'} \frac{\delta(\target-z)\delta(\omega_1+\omega')}{-i\omega_1+\eps} T(\start,z,\omega')\deltabar(\omega'+\omega_0) \nonumber \\
	&= \carryingCap^{-1} \gamma \tau \frac{\trans{\start,\target,-\omega_1}}{-i\omega_1+\eps}\deltabar(\omega_0-\omega_1),
\end{align}
and to second order,
\begin{align}
&- \carryingCap^{-1} \gamma^2 \lambda \sigma \iint \dd{z_1}\dd{z_2}\dbar{\omega'_1}\dbar{\omega'_2}\dbar{\omega{''}_1}\dbar{\omega_2{''}} \\
	&\qquad \times\avg{\psi(\target,\omega_1) \tpsi(z_1,-\omega'_1-\omega''_1)\psi(z_1,\omega_1{''})\phi(z_1,\omega'_1)\tpsi(z_2,\omega_2{''}) \tphi(z_2,\omega'_2)\phi(z_2,-\omega'_2-\omega''_2)\tphi(\start,\omega_0)  }_{\cS;\deprate=0} \nonumber \\
	&\qquad \qquad = -\carryingCap^{-1} \gamma^2 \lambda \sigma \frac{1}{-i\omega_1 +\eps} \int \dbar{\omega''_2} \frac{\return{\target,\target,\omega_1-\omega''_2}}{-i\omega''_1 + \eps} \trans{\start,\target,\omega_0}\deltabar(\omega_0 + \omega_1)
\\
	&\qquad \qquad = -\carryingCap^{-1} \gamma^2 \lambda \sigma \return{\target,\omega_1+i\eps} \frac{\trans{\start,\target ,\omega_0}}{-i\omega_1 + \varepsilon}  \deltabar(\omega_0 + \omega_1),
	\label{}
\end{align}
where in the first equality we used the definition of the return probability to abbreviate
\begin{align}
	\iint \dd{z_1}\dd{z_2} \deltabar(\target-z_1)\delta(z_1-z_2)\trans{z_1,z_2,\omega} = \return{\target,\omega},
	\label{}
\end{align}
and in the second equality used Cauchy's residue formula to solve the integral by evaluating the residue of the simple pole at $\omega''_2 = -i\eps$. 

Since both correlators in \Eqref{app:eq:wick_sum} are proportional to $\deltabar(\omega_0 + \omega_1)$, all higher order expansion terms  factorise, after integrating over the internal frequencies, into a product over amputated one-loop bubble diagrams (\ie interpreted here as a function of external parameters $z_1,\omega_1$ and $z_2,\omega_2$, respectively) which by analogous reasoning to the calculation above evaluate as
\begin{align}
	\tikz[baseline=-5pt]{
    \draw[Aactivity] (0,0) -- (.5,0) ;
    \draw[Bsubstrate] (0.0,0) arc (180:360:.25);
    \node at (-0.7,0) {$z_1,\omega_1$};
    \node at (1.2,0) {$z_2,\omega_2$};
    } \ 
    & = \gamma^2 \lambda \sigma\return{z_1,\omega_1+i\eps} \delta(\omega_1+\omega_2)\delta(z_1-z_2).
	\label{appeq:bubble_is_R}
\end{align}
The bubble diagram graphically encodes the probability of a particle depositing a trace and then returning to it, in other words a ``time-ordered'' return probability: The green wriggly line may be understood as the trace which once placed remains immobile, meanwhile the red solid line represents the diffusing, and returning, walker. Likewise, the higher order diagrams in \Eqref{eq:tau_r_diagram_sum} may be interpreted as repeated returns to $\target$. 

Returning to \Eqref{eq:tau_r_diagram_sum}, the renormalised $\tau$ coupling, $\tau_R$, is the effective factor satisfying
\begin{align}
	\avg{\psi(\target,\omega_1)\tphi(\start,\omega_0)}_{\cS} = \frac{1}{-i\omega_1 + \varepsilon}  \tau_R(\omega_1)  \trans{\start,\target,-\omega_0}  \deltabar(\omega_0 + \omega_1)		\label{}
\end{align}
and collecting the factors generated by the terms in \Eqref{appeq:path_avg_1}, and evaluated using \Eqref{appeq:bubble_is_R}, one obtains
\begin{align}
	\tau_R(\omega_1) &= \gamma \tikz[baseline=-2.5pt]{
    \node at (0,.3) {$\tau$};
    \draw[Aactivity] (0,0) -- (.2,0) ;
    \draw[Bsubstrate] (-.2,0) -- (0,0); 
    }
    + \gamma^2
    \tikz[baseline=-2.5pt]{
    \node at (0,.3) {$-\lambda$};
    \node at (0.5,.3) {$\sigma$};
    \draw[Bsubstrate] (-0.3,0) -- (-0,0) ;
    \draw[Aactivity] (0,0) -- (.8,0) ;
    \draw[Bsubstrate] (0.0,0) arc (180:360:.25);
    }
     + \gamma^3
    \tikz[baseline=-2.5pt]{
        \node at (0,.3) {$-\lambda$};
    \node at (0.5,.3) {$-\kappa$};
    \node at (1.0,.3) {$\sigma$};
    \draw[Bsubstrate] (-0.3,0) -- (-0,0) ;
    \draw[Aactivity] (0,0) -- (1.3,0) ;
    \draw[Bsubstrate] (0.0,0) arc (180:360:.25);
    \draw[Bsubstrate] (0.5,0) arc (180:360:.25);
    }+ ...\\
    &= \left[\gamma \tau - \gamma^2 \lambda \sigma \return{\target,\omega_1 + i \eps} + \gamma^3 \lambda \sigma \kappa \left( \return{\target,\omega_1 + i \eps} \right)^2 - \gamma^4 c \sigma \kappa^2 \left( \return{\target,\omega_1 + i \eps} \right)^3 + \cdots\right]\deltabar(\omega_0 + \omega_1)
\end{align}
This series can be resummed using the geometric series, in field theory often referred to as Dyson summation \cite{LeBellac:1991}. Rearranging the sum gives
\begin{align}
	\tau_R(\omega_1) &= \left[ \gamma \tau + \gamma \frac{\lambda \sigma}{\kappa} \sum_{r=1}^{\infty} \left (-\gamma \kappa \return{\target,\omega_1+i\eps} \right)^r \right]\\
	&= \left[ \gamma \tau + \gamma \frac{\lambda \sigma }{\kappa}\sum_{r=0}^{\infty} \left (-\gamma \kappa \return{\target,\omega_1+i\eps} \right)^r - \gamma \frac{c\sigma}{\kappa}\right]\\
	&= \left[
	\frac{\gamma\tau}{1 + \gamma \kappa \return{\target,\omega_1+i\eps}} + \gamma\left(\underbrace{ \tau - \frac{c\sigma}{\kappa} }_{=0}  \right)\right]\\
	& = \frac{\gamma \tau}{1 + \gamma  \kappa \return{\target,\omega_1+i\eps}}
	\label{appeq:geom_sum_loop}
\end{align}
where we made use of the bare values given in \Eqref{eq:sdep_def}, \ie replaced $\sigma$ with $\tau$ and used $\lambda = \kappa$ at bare level. This vertex  interpolates the physical pictures for $\gamma = 0$, where no deposition takes place ($\tau_R = 0$), and $\gamma \to \infty$, where every newly visited site gets marked immediately by a deposited trace. For $\gamma \to \infty$, the coupling tends to
\begin{align}
	\lim_{\gamma \to \infty} \tau_R(\omega_1) = \frac{\carryingCap
}{\return{\target,\omega_1+i\eps}},
	\label{}
\end{align}
using $\carryingCap=\tau/\kappa$.

\section{Derivation of four non-Markovian correction terms to trace function\label{app:non_markovian_Q}}
When expanding the nonlinear action in Eq.~\eqref{eq:trace_exp_in_g2} to all orders in $\gamma$ and to $\cO(\cpl^2)$, the nonvanishing contribution of joint perturbative order $\gamma^n \cpl^2$ is obtained by inserting the (Fourier-transformed) $\cpl^2$-decoration
\begin{align}
&\tikz[baseline=-2.5pt]
{
	\draw[Aactivity] (0,0) -- (1,0);
	\draw[Aactivity] (2,0) -- (3,0);
	\draw[Bactivity] (2.5,0) arc (0:180:1.);
	\draw[very thick] (.6,-0.1) -- (0.6,0.1);
	\draw[very thick] (2.6,-0.1) -- (2.6,0.1);
	\node at (0,0) [anchor=east] {$y_1,\tw_1$};
	\node at (1,-.1) [anchor=north] {$y_2,\tw_2$};
	\node at (2,-.1) [anchor=north] {$y_3,\tw_3$};
	\node at (3,0.) [anchor=west] {$y_4,\tw_4$};
	} \\
	\nonumber
	&= \iint \ddbar{\tw_1} \ldots \ddbar{\tw_4} \dint{y_1} ...\dint{y_4} \tphi(y_1,\tw_1) \partial_{y_2} \phi(y_2,\tw_2) C_2(\tw_3 + \tw_4) \tphi(y_3,\tw_3) \partial_{y_4} \phi(y_4,\tw_4) \times \deltabar(\tw_1+\tw_2+\tw_3+\tw_4)
\end{align}
into the expansion of the trace function $\avg{\psi \tphi}$ as
\begin{align}
\label{appeq:Q_expansion_g2}
    & \!\!\!\!\!\!\!\!\!\!\!\trace(\start,\target,\omega) \deltabar(\omega + \omega') \\
    \nonumber
     =& \sum_{n=0}^{\infty} \gamma^{2+n} \int \dbar{\omega_1} ... \dbar{\omega_n} \dbar{\tw_1} ... \dbar{\tw_4} \dint{z_1} ... \dint{z_n} \dint{y_1} ...\dint{y_4}  \\
    &
    \left \langle 
    \psi(\target, \omega) \left (-\lambda \tpsi(z_1,\omega_1) \phi(z_1,\omega_1) \psi(z_1,\omega_1) \right)
    \left (-\kappa \tphi(z_2,\omega_2) \tpsi(z_2,\omega_2) \phi(z_2,\omega_2) \psi(z_2,\omega_2) \right) \right. \nonumber \\
    &...
   \left( -\kappa \tphi(z_{n-1},\omega_{n-1}) \tpsi(z_{n-1},\omega_{n-1}) \phi(z_{n-1},\omega_{n-1}) \psi(z_{n-1},\omega_{n-1}) \right)
    \left (\sigma \tphi(z_n,\omega_n) \tpsi(z_{n},\omega_{n}) \phi(z_{n},\omega_{n}) \right) 
    \phi(\start,\omega') \nonumber
  \\
  &\times \left. \cpl^2 \tphi(y_1,\tw_1) \partial_{y_2} \phi(y_2,\tw_2) C_2(\tw_3 + \tw_4) \tphi(y_3,\tw_3) \partial_{y_4} \phi(y_4,\tw_4) \times \deltabar(\tw_1+\tw_2+\tw_3+\tw_4) \right \rangle. \nonumber
\end{align}

As usual, we employ Wick's theorem to evaluate this average over Gaussian random variables: When expanding the average $\carryingCap^{-1} \avg{ \psi(\target, \omega) \tphi(\start, \omega')}$ in powers of $\gamma$ and $\cpl$, as shown in \Eqref{appeq:Q_expansion_g2}, the resulting coefficients are averages over finite products of $\phi_1, ...\phi_{j_1}, \tphi_1... \phi_{j_2}, \psi_1 ... \psi_{j_3}$ and $\tpsi_1 .. \tpsi_{j_4}$ fields (where $\phi_i = \phi(z_i, \omega_i)$ etc.) with respect to the Gaussian measure (defined by the action in \Eqref{eq:joint_field_measure} for $\gamma = 0$). Each of those coefficients is evaluated using Wick's theorem, \ie 
\begin{align}
    \avg{ \psi_{1} ... \psi_{j_1} \tpsi_{1} ... \tpsi_{j_2} \phi_{1} ... \phi_{j_3} \tphi_1 ... \tphi_{j_4}}_{\cS; \gamma = 0} = \sum_{\text{pairings}} \prod_{(k_m, \ell_m)} \avg{ \upphi_{k_m} \upphi_{\ell_m}}_{\cS; \gamma = 0}
    \label{appeq:wick}
\end{align}
where the sum runs over all possible pairwise pairings of the $j_1 + j_2 + j_3 + j_4$ indices, and we use $\upphi$ in lieu of $\phi, \tphi, \psi, \tpsi$ to alleviate notation. Although the number of possible combinations of such pairings is very large, the right hand side of \Eqref{appeq:wick} drastically simplifies because most of the pairwise averages vanish under the Gaussian average.
Again, as in the case of $\cpl=0$ (cf.~App.~\ref{app:markovian_visit_p}), the only non-vanishing Gaussian correlators are those of the form $\avg{\phi \tphi},\avg{\psi \tpsi}$ as given by Eqs.~\eqref{app:eq:wick_sum}, \eqref{appeq:Psipsi_propagator}.
Thus, the sum over averages in \Eqref{appeq:Q_expansion_g2} simplifies into a sum over integrals over products of these two Gaussian propagators. The integrals run over $n$ internal spatial variables $z_1, ..., z_n$ which stem from the expansion in $\gamma^n$, and four internal spatial variables $y_1,...,y_4$ which stem from the expansion in $\cpl^2$ (and thus are related to the non-Markovian correction). Analagously, the integration also runs over $n$ frequencies $\omega_1,...,\omega_n$ stemming from the $\gamma^n$-expansion and four frequencies $\tw_1, ...,\tw_4$ from the $\cpl^2$-expansion. 
The integral over the internal space variables $z_1,...,z_n$, simplifies significantly, since the corresponding $\avg{\psi\tpsi}$ propagators (cf.~  \Eqref{appeq:Psipsi_propagator})  are proportional to spatial $\delta$-functions. 
This results in all internal space coordinates to identify as $z_1 = ... = z_n = \target$. 

The integral over the internal spatial coordinates $y_1,...,y_4$ appearing in the four $\phi$ and $\tphi$ fields in \Eqref{appeq:Q_expansion_g2}, however, is less trivial as it involves the correlators (``propagators'') of $\avg{\phi \tphi}$. Likewise, the integration over $\tw_1, ..., \tw_4$, further involves the correlator of the driving noise, $\corr{\tw_3 + \tw_4}$. Hence, the integrals stemming from the $\cpl^2$-expansion need to be dealt with more carefully:
The four fields $\tphi(y_1, \tw_1) \partial_{y_2} \tphi(y_2, \tw_2) \tphi(y_3, \tw_4) \partial_{y_4} \phi(y_4, \tw_4)$ appearing in every term of \Eqref{appeq:Q_expansion_g2} have to be paired up (according to \Eqref{appeq:wick}) with another creation field $\tphi(z_i)$ or annihilation field $\tphi(z_i)$ appearing in \Eqref{appeq:Q_expansion_g2}, respectively, in order to have a non-vanishing contribution (\cf \Eqref{app:eq:wick_sum}). Up to permutation of indices,  each of these possible combinations (which we refer to as ``\emph{Wick pairings}'', \Eqref{appeq:wick}) can fundamentally be grouped into four different ways that are best understood diagrammatically:
In any case, each of the two $\phi(y_i) \partial_{y_j} \tphi(y_j)$ attaches via Wick-pairing to a $\phi(z_k) \tphi(z_{\ell})$  field appearing in the terms of \Eqref{appeq:Q_expansion_g2}. Concurrently, $\tphi(z_k), \phi(z_{\ell})$,  Wick-pair with two other corresponding fields giving rise to connected $\avg{\phi \tphi}$ - ``propagators'' and so on. Thereby, the Wick-pairing of $\phi(y_i) \partial_{y_j} \tphi(y_j)$, diagrammatically speaking, splits an existing propagator in the $\gamma^n$-expansion into two or, as shown below, in three.

In the expansion to order $\gamma^n$, there is one propagator from $\start$ to $\target$ (\emph{transition}) and $n$ propagators from $\target$ to $\target$ (\emph{return}) represented by loop diagrams. The $\cpl$-vertex can thus occur in four different ways, Fig.~\ref{fig:four_diagrams}, to be distinguished by whether and how the $\cpl$-vertex enters into the initial \emph{transition propagator}, $\tikz{\draw[Aactivity] (-.35,0) -- (.35,0)}$, 
in 
$\tikz{\draw[Aactivity] (0,0) -- (.4,0) ; 
    \draw[Bsubstrate] (-.4,0) -- (.,0) ;} +
 \tikz{\draw[Bsubstrate] (-0.3,0) -- (-0,0) ;
    \draw[Aactivity] (0,0) -- (.6,0) ;
    \draw[Bsubstrate] (0.0,0) arc (180:360:.17);}
    + \ldots$, \Eqref{tau_R_diagrams}, or into any of the \emph{return propagators}
    $\tikz{\draw[Aactivity] (0,0) -- (.34,0) ;
    \draw[Bsubstrate] (0.0,0) arc (180:360:.17);}$.
First, we consider $\traceI$, I) in Fig.~\ref{fig:four_diagrams}, the case of the correlation coupling appearing twice within the same return propagator leading to the diagrammatic sum

\begin{align}\label{appeq:Q1_diagrammatic_sum}
\traceI(\start,\target,\omega) &=\carryingCap^{-1} \lim_{\gamma \to \infty} \left[ -\gamma^2 \tikz[baseline=-5pt]{
    \draw[Aactivity] (-.5,0) -- (1.,0) ;
    \draw[Bsubstrate] (-1,0) -- (-.5,0);
    \draw[Bsubstrate] (-.5,0) arc(180:360:.5);
    \draw[Bactivity] (.25,0) arc (0:180:.25);
	\draw[very thick] (-.15,-0.1) -- (-.15,0.1);
	\draw[very thick] (.35,-0.1) -- (.35,0.1);
   } +
   \gamma^3\tikz[baseline=-5pt]{
    \draw[Aactivity] (-.5,0) -- (1.5,0) ;
    \draw[Bsubstrate] (-1,0) -- (-.5,0);
    \draw[Bsubstrate] (-.5,0) arc(180:360:.5);
    \draw[Bsubstrate] (0.5,0) arc(180:360:.25);
    \draw[Bactivity] (.25,0) arc (0:180:.25);
	\draw[very thick] (-.15,-0.1) -- (-.15,0.1);
	\draw[very thick] (.35,-0.1) -- (.35,0.1);
   }+
   \gamma^3\tikz[baseline=-5pt]{
    \draw[Aactivity] (-1.,0) -- (1.,0) ;
    \draw[Bsubstrate] (-1.5,0) -- (-1.,0);
    \draw[Bsubstrate] (-.5,0) arc(180:360:.5);
    \draw[Bsubstrate] (-1.,0) arc(180:360:.25);
    \draw[Bactivity] (.25,0) arc (0:180:.25);
	\draw[very thick] (-.15,-0.1) -- (-.15,0.1);
	\draw[very thick] (.35,-0.1) -- (.35,0.1);
   }+... \right]
   \\
   &\label{Q1_with_geoSums}
   = \carryingCap^{-1} \lim_{\gamma \to \infty} \left[ \gamma^2 \tikz[baseline=-5pt]{
    \draw[Bsubstrate] (-.5,0) -- (0,0);
    }
    	\sum_{r=0}^{\infty} \left( \gamma \tikz[baseline=-5pt]{
    \draw[Aactivity] (-.5,0) -- (0,0) ;
    \draw[Bsubstrate] (-.5,0) arc(180:360:.25);
   } \right)^r
   \times
   \tikz[baseline=-5pt]{
    \draw[Aactivity] (0,0) -- (1,0);
	\draw[Bactivity] (.75,0) arc(0:180:.25);
	\draw[Bsubstrate] (0,0) arc(180:360:.5);
	\draw[very thick] (.85,-0.1) -- (0.85,0.1);
	\draw[very thick] (.35,-0.1) -- (.35,0.1);   
	}
	\times
		\sum_{s=0}^{\infty} 
		\left( \gamma \tikz[baseline=-5pt]{
    \draw[Aactivity] (-.5,0) -- (0,0) ;
    \draw[Bsubstrate] (-.5,0) arc(180:360:.25);   } \right)^{s}
    \times
\tikz[baseline=-5pt]{
    \draw[Aactivity] (0,0) -- (0.5,0); 
} 
    \right]\\
&=   -\frac{1}{-i \omega} \frac{\trans{\start,\target}}{\left(\return{\target,\omega}\right)^2} \iint \dint{y_1} \dint{y_2} \dbar{\tw} \left( \partial_{y_1} \trans{\target,y_1,\omega} \right) \left(\partial_{y_2} \trans{y_1,y_2,\omega-\tw}\right) \trans{y_2,\target,\omega} \corr{\tw}
\label{appeq:Q1_as_integral}
\end{align}

In the last line, we replaced the geometric sums in \Eqref{Q1_with_geoSums} by the expression found for $\tau_R$ in \Eqref{appeq:geom_sum_loop}. To be precise, this is not a matter of trivially renormalising $\tau$ to $\tau_R$ , as the sums appearing in \Eqref{Q1_with_geoSums} are in fact renormalisations of $\lambda$ and $\sigma$, respectively,
\begin{align}
\label{appeq:geom_sum_loop_lambda}
-\lambda_R &= (-\lambda) \gamma\sum_{r=0}^{\infty} 
	\left( 
	\gamma \tikz[baseline=-5pt]{
    \draw[Aactivity] (-.5,0) -- (0,0) ;
    \draw[Bsubstrate] (-.5,0) arc(180:360:.25);
   }
    \right)^r
     =  \gamma (-\lambda) + \gamma (-\lambda)
     \sum_{r=1}^{\infty}  
     \left( 
     -\gamma \kappa \return{\target,\omega} \right)^r \stackrel{\gamma \to \infty}	{\longrightarrow} - \frac{1}{\return{\target,\omega}} \\
\sigma_R &= \sigma \gamma\sum_{r=0}^{\infty} 
	\left( 
	\gamma \tikz[baseline=-5pt]{
    \draw[Aactivity] (-.5,0) -- (0,0) ;
    \draw[Bsubstrate] (-.5,0) arc(180:360:.25);
   }
    \right)^r
     =  \gamma \sigma + \gamma \sigma
     \sum_{r=1}^{\infty}  
     \left( 
     -\gamma \kappa \return{\target,\omega} \right)^r \stackrel{\gamma \to \infty}	{\longrightarrow} \frac{\carryingCap}{\return{\target,\omega}},
\end{align}
where again we made use of the definition $\carryingCap = \sigma \kappa^{-1}$.
However, comparing to \Eqref{appeq:geom_sum_loop}, they renormalise identically, as does $\kappa$
\begin{align}
\label{appeq:geom_sum_loop_kappa}
    -\kappa_R &= (-\kappa) \gamma\sum_{r=0}^{\infty} 
	\left( 
	\gamma \tikz[baseline=-5pt]{
    \draw[Aactivity] (-.5,0) -- (0,0) ;
    \draw[Bsubstrate] (-.5,0) arc(180:360:.25);
   }
    \right)^r
     =  \gamma(-\kappa)
     \sum_{r=0}^{\infty}  
     \left( 
     -\gamma \kappa \return{\target,\omega} \right)^r \stackrel{\gamma \to \infty}	{\longrightarrow} - \frac{1}{\return{\target,\omega}} 
\end{align}
to be used below.

Another identity entering into expressing $\traceI$, \Eqref{appeq:Q1_diagrammatic_sum}, as \Eqref{appeq:Q1_as_integral}, is the key-ingredient of $\traceI$,
\begin{align}\label{appeq:central_vertex_traceI}
\tikz[baseline=-5pt]{
 	\draw (0,0) node[anchor=east] {$(\target,\omega)$};
	\draw (1.5,0) node[anchor=west] {$(\target,\omega)$};
    \draw[Aactivity] (0,0) -- (1.5,0);
	\draw[Bactivity] (1.25,0) arc(0:180:.5) node[pos=.5,anchor=south] {$\tw$};
	\draw[Bsubstrate] (0,0) arc(180:360:.75);
	\draw[very thick] (1.35,-0.1) -- (1.35,0.1) node [above=1mm] {\tiny$\partial_{y_1}$};
	\draw[very thick] (.35,-0.1) -- (.35,0.1) node [below=2mm] {\tiny$\partial_{y_2}$};   
	} &= \iint \dint{y_1} \dint{y_2} \dbar{\tw} \left( \partial_{y_1} \trans{\target,y_1,\omega} \right) \left(\partial_{y_2} \trans{y_1,y_2,\omega-\tw}\right) \trans{y_2,\target,\omega} \corr{\tw}
\end{align}

Considering $\traceII$, shown as II) in Fig.~\ref{fig:four_diagrams}, next, the two $\cpl$-vertices may be inserted into two different return propagators 
$\tikz{\draw[Aactivity] (0,0) -- (.34,0) ;
    \draw[Bsubstrate] (0.0,0) arc (180:360:.17);}$
of a diagram in the expansion \Eqref{tau_R_diagrams},
\begin{align}
\traceII(\start,\target,\omega) &= \carryingCap^{-1} \lim_{\gamma \to \infty} \left[
\gamma^3 
\tikz[baseline=-5pt]
{
\draw[Bsubstrate] (-0.5,0) -- (0,0);
\draw[Aactivity] (0,0) -- (1.5,0);
\draw[Bsubstrate] (0,0) arc(180:360:.25);
\draw[Bsubstrate] (0.5,0) arc(180:360:.25);
\draw[very thick] (.85,-0.1) -- (.85,0.1);
\draw[very thick] (.35,-0.1) -- (.35,0.1);
\draw[Bactivity] (.75,0) arc (0:180:.25);
}
-
\gamma^4 \tikz[baseline=-5pt]{
\draw[Bsubstrate] (-1.,0) -- (-0.5,0);
\draw[Aactivity] (-0.5,0) -- (1.5,0);
\draw[Bsubstrate] (-0.5,0) arc(180:360:.25);
\draw[Bsubstrate] (0,0) arc(180:360:.25);
\draw[Bsubstrate] (0.5,0) arc(180:360:.25);
\draw[very thick] (.85,-0.1) -- (.85,0.1);
\draw[very thick] (.35,-0.1) -- (.35,0.1);
\draw[Bactivity] (.75,0) arc (0:180:.25);
}
-
\gamma^4 \tikz[baseline=-5pt]{
\draw[Bsubstrate] (-.5,0) -- (-0.,0);
\draw[Aactivity] (0,0) -- (2.,0);
\draw[Bsubstrate] (1.,0) arc(180:360:.25);
\draw[Bsubstrate] (0,0) arc(180:360:.25);
\draw[Bsubstrate] (0.5,0) arc(180:360:.25);
\draw[very thick] (1.35,-0.1) -- (1.35,0.1);
\draw[very thick] (.35,-0.1) -- (.35,0.1);
\draw[Bactivity] (1.25,0) arc (0:180:.5);
}
-
\gamma^4 \tikz[baseline=-5pt]{
\draw[Bsubstrate] (-.5,0) -- (-0.,0);
\draw[Aactivity] (0,0) -- (2.,0);
\draw[Bsubstrate] (1.,0) arc(180:360:.25);
\draw[Bsubstrate] (0,0) arc(180:360:.25);
\draw[Bsubstrate] (0.5,0) arc(180:360:.25);
\draw[very thick] (.85,-0.1) -- (.85,0.1);
\draw[very thick] (.35,-0.1) -- (.35,0.1);
\draw[Bactivity] (.75,0) arc (0:180:.25);
}+... \right]
\\
&=\carryingCap^{-1} \lim_{\gamma \to \infty} \left[ \gamma^3 \tikz[baseline=-5pt]{
    \draw[Bsubstrate] (-.5,0) -- (0,0);
    }
    	\sum_{r=0}^{\infty} \left( \gamma \tikz[baseline=-5pt]{
    \draw[Aactivity] (-.5,0) -- (0,0) ;
    \draw[Bsubstrate] (-.5,0) arc(180:360:.25);
   } \right)^r
   \times
   \tikz[baseline=-5pt]{
    \draw[Aactivity] (0,0) -- (.5,0);
	\draw[Bsubstrate] (0,0) arc(180:360:.25);
	\draw[very thick] (.35,-0.1) -- (.35,0.1);   
	\draw (1.3,0) node {$\times \sum_{s=0}^{\infty}\! \big(\gamma$};
    \draw[Aactivity] (2.1,0) -- (2.6,0) ;
    \draw[Bsubstrate] (2.1,0) arc(180:360:.25);  
    \draw (2.9,0) node {$\big)^{s} \times$}; 
    \draw[Aactivity] (3.2,0) -- (3.7,0);
    \draw[Bsubstrate] (3.2,0) arc(180:360:.25);
    \draw[Bactivity] (3.5,0) arc (60:120:3.3);
    \draw[very thick] (3.6,-0.1) -- (3.6,0.1); 
    }
    \times
    \sum_{t=0}^{\infty} \left( \gamma \tikz[baseline=-5pt]{
    \draw[Aactivity] (-.5,0) -- (0,0) ;
    \draw[Bsubstrate] (-.5,0) arc(180:360:.25);
   } \right)^t 
   \times
   \tikz[baseline=-5pt]{
    \draw[Aactivity] (-.5,0) -- (0,0) ;}
    \right]
    \label{appeq:Q2_as_geometric_sum} 
     \\
   &= \carryingCap^{-1} \lim_{\gamma \to \infty} \left[ \gamma^3  \frac{1}{-i \omega} \frac{- \lambda \gamma}{1 + \gamma \kappa R(\target, \omega)}
   \times \tikz[baseline=-5pt]{
    \draw[Aactivity] (0,0) -- (.5,0);
	\draw[Bsubstrate] (0,0) arc(180:360:.25);
	\draw[very thick] (.35,-0.1) -- (.35,0.1);   
	\draw (1.3,0) node {$\times \sum_{s=0}^{\infty}\! \big(\gamma$};
    \draw[Aactivity] (2.1,0) -- (2.6,0) ;
    \draw[Bsubstrate] (2.1,0) arc(180:360:.25);  
    \draw (2.9,0) node {$\big)^{s} \times$}; 
    \draw[Aactivity] (3.2,0) -- (3.7,0);
    \draw[Bsubstrate] (3.2,0) arc(180:360:.25);
    \draw[Bactivity] (3.5,0) arc (60:120:3.3);
    \draw[very thick] (3.6,-0.1) -- (3.6,0.1); 
    }
    \times \frac{ \sigma \gamma}{1 + \gamma \kappa R(\target, \omega)} T(\start, \target, \omega) \right]
    \label{appeq:Q2_intermediate_result}
\end{align}

Here, we made use again of the geometric sums in 
\Eqref{appeq:geom_sum_loop} as well as
Eqs.~\eqref{appeq:geom_sum_loop_lambda}--\eqref{appeq:geom_sum_loop_kappa}, 
which features three times in \Eqref{appeq:Q2_as_geometric_sum}, once with dummy index $r$, once with $s$ and once with $t$. The central one, running with index $s$ differs from the others by the (blue) dashed line that represents the noise carrying momentum $\tw$ thus bypassing the loops, so that only $\omega-\tw$ flows through loops summed over. Using \Eqref{appeq:geom_sum_loop_kappa} in this loop, the effective vertex of $\traceII(\start,\target,\omega)$  is
\begin{align}
\label{appeq:enclosed_Q_diagram}
   &\tikz[baseline=-5pt]{
    \draw[Aactivity] (-.5,0) -- (.5,0);
	\draw[Bsubstrate] (-.5,0) arc(180:360:.5);
	\draw[very thick] (.15,-0.1) -- (.15,0.1) node[above=1mm] {\tiny$\partial_{y_2}$};   
	\draw (1.3,-0.1) node {$\times \sum_{s=0}^{\infty}\! \big(\gamma$};
    \draw[Aactivity] (2.1,-0.1) -- (2.6,-0.1) ;
    \draw[Bsubstrate] (2.1,-0.1) arc(180:360:.25);  
    \draw (2.9,-0.1) node {$\big)^{s} \times$}; 
    \draw[Aactivity] (3.2,0) -- (4.2,0);
    \draw[Bsubstrate] (3.2,0) arc(180:360:.5);
    \draw[Bactivity] (3.7,0) arc (60:120:3.7);
    \draw[very thick] (3.8,-0.1) -- (3.8,0.1) node[below=1mm] {\tiny$\partial_{y_1}$}; 
    \draw (4.9,0) node {$(\target,\omega)$};
    \draw (-1.2,0) node {$(\target,\omega)$};
    \draw (2.35,.9) node {$\tw$};
    \draw[->,>=latex] (2.7,-0.7) -- (.7,-0.7) node[below,pos=0.5] {$\omega-\tw$};
    } \\
    &= 
    \iint \dint{y_1} \dint{y_2} \dbar{\tw} 
    \left( \partial_{y_1} \trans{\target,y_1,\omega} \right) \trans{y_1,\target,\omega-\tw} \frac{1}{\return{\target,\omega-\tw}} \left(\partial_{y_2} \trans{\target,y_2,\omega-\tw}\right) \trans{y_2,\target,\omega} \corr{\tw} \ ,
    \label{appeq:overlapping_loop}
\end{align}
to be contrasted with \Eqref{appeq:central_vertex_traceI}, the effective vertex of $\traceI(\start,\target,\omega)$.

Inserting the result \eqref{appeq:overlapping_loop} into Eq.~\eqref{appeq:Q2_intermediate_result}, and using $\carryingCap = \sigma/ \kappa$, we obtain an explicit formula
\begin{multline}
\traceII(\start, \target, \omega) \\
 = \label{appeq:Q2_as_integral} \frac{1}{-i \omega} \frac{\trans{\start,\target}}{\left(\return{\target,\omega}\right)^2} \iint \dint{y_1} \dint{y_2} \dbar{\tw} \frac{\left( \partial_{y_1} \trans{\target,y_1,\omega} \right) \trans{y_1,\target,\omega-\tw}}{\return{\target,\omega-\tw}} \left(\partial_{y_2} \trans{\target,y_2,\omega-\tw}\right) \trans{y_2,\target,\omega} \corr{\tw}
\end{multline}

Thirdly, we consider 
$\traceIII$, shown as III) in Fig.~\ref{fig:four_diagrams},
the case of the transition propagator, 
$\tikz{\draw[Aactivity] (-.35,0) -- (.35,0)}$,
coupling to one of the return propagators, 
$\tikz{\draw[Aactivity] (0,0) -- (.34,0) ;
    \draw[Bsubstrate] (0.0,0) arc (180:360:.17);}$,
via $\corr{\omega}$ which results in the diagrammatic expansion
\begin{align}
 \traceIII(\start,\target,\omega) &= \carryingCap^{-1} \lim_{\gamma \to \infty} \left[-
\gamma^2 
\tikz[baseline=-5pt]
{
\draw[Bsubstrate] (-0.5,0) -- (0,0);
\draw[Aactivity] (0,0) -- (1.,0);
\draw[Bsubstrate] (0,0) arc(180:360:.25);
\draw[very thick] (.85,-0.1) -- (.85,0.1);
\draw[very thick] (.35,-0.1) -- (.35,0.1);
\draw[Bactivity] (.75,0) arc (0:180:.25);
}
+
\gamma^3 \tikz[baseline=-5pt]{
\draw[Bsubstrate] (-1.,0) -- (-0.5,0);
\draw[Aactivity] (-0.5,0) -- (1.0,0);
\draw[Bsubstrate] (-0.5,0) arc(180:360:.25);
\draw[Bsubstrate] (0,0) arc(180:360:.25);
\draw[very thick] (.85,-0.1) -- (.85,0.1);
\draw[very thick] (.35,-0.1) -- (.35,0.1);
\draw[Bactivity] (.75,0) arc (0:180:.25);
}
+\gamma^3 \tikz[baseline=-5pt]{
\draw[Bsubstrate] (-1.,0) -- (-0.5,0);
\draw[Aactivity] (-0.5,0) -- (1.,0);
\draw[Bsubstrate] (-0.5,0) arc(180:360:.25);
\draw[Bsubstrate] (0,0) arc(180:360:.25);
\draw[very thick] (.85,-0.1) -- (.85,0.1);
\draw[very thick] (-.15,-0.1) -- (-.15,0.1);
\draw[Bactivity] (.75,0) arc (50:130:.8);
}
+... \right]
\\
&= \carryingCap^{-1} \lim_{\gamma \to \infty} \left[ \gamma^3 \tikz[baseline=-5pt]{
    \draw[Bsubstrate] (-.5,0) -- (0,0);
    }
    	\sum_{r=0}^{\infty} \left( \gamma \tikz[baseline=-5pt]{
    \draw[Aactivity] (-.5,0) -- (0,0) ;
    \draw[Bsubstrate] (-.5,0) arc(180:360:.25);
   } \right)^r
   \times
   \tikz[baseline=-5pt]{
    \draw[Aactivity] (0,0) -- (.5,0);
	\draw[Bsubstrate] (0,0) arc(180:360:.25);
	\draw[very thick] (.35,-0.1) -- (.35,0.1);   
	\draw (1.3,0) node {$\times \sum_{s=0}^{\infty}\! \big(\gamma$};
    \draw[Aactivity] (2.1,0) -- (2.6,0) ;
    \draw[Bsubstrate] (2.1,0) arc(180:360:.25);  
    \draw (2.9,0) node {$\big)^{s} \times$}; 
    \draw[Aactivity] (3.2,0) -- (3.9,0);
    \draw[Bactivity] (3.5,0) arc (60:120:3.3);
    \draw[very thick] (3.6,-0.1) -- (3.6,0.1); 
    } \right]\\
    & = \label{appeq:QIII_final_loop} \carryingCap^{-1} \lim_{\gamma \to \infty} \left[ \gamma^3 \frac{1}{-i \omega} \frac{(-\lambda) \gamma}{1 + \gamma \kappa \return{\target, \omega}} \times
   \tikz[baseline=-5pt]{
    \draw[Aactivity] (0,0) -- (.5,0);
	\draw[Bsubstrate] (0,0) arc(180:360:.25);
	\draw[very thick] (.35,-0.1) -- (.35,0.1);   
	\draw (1.3,0) node {$\times \sum_{s=0}^{\infty}\! \big(\gamma$};
    \draw[Aactivity] (2.1,0) -- (2.6,0) ;
    \draw[Bsubstrate] (2.1,0) arc(180:360:.25);  
    \draw (2.9,0) node {$\big)^{s} \times$}; 
    \draw[Aactivity] (3.2,0) -- (3.9,0);
    \draw[Bactivity] (3.5,0) arc (60:120:3.3);
    \draw[very thick] (3.6,-0.1) -- (3.6,0.1); 
    } \right]
\end{align}
where we made use of the renormalisation of $\lambda$, \Eqref{appeq:geom_sum_loop_lambda}.
For the remaining diagram in \Eqref{appeq:QIII_final_loop}, which differs from \Eqref{appeq:overlapping_loop} only by an incoming transition propagator instead of a return propagator, we find
\begin{align}
   &\tikz[baseline=-5pt]{
    \draw[Aactivity] (0,0) -- (.5,0);
	\draw[Bsubstrate] (0,0) arc(180:360:.25);
	\draw[very thick] (.35,-0.1) -- (.35,0.1);   
	\draw (1.3,0) node {$\times \sum_{s=0}^{\infty}\! \big(\gamma$};
    \draw[Aactivity] (2.1,0) -- (2.6,0) ;
    \draw[Bsubstrate] (2.1,0) arc(180:360:.25);  
    \draw (2.9,0) node {$\big)^{s} \times$}; 
    \draw[Aactivity] (3.2,0) -- (3.9,0);
    \draw[Bactivity] (3.5,0) arc (60:120:3.3);
    \draw[very thick] (3.6,-0.1) -- (3.6,0.1); 
    } \nonumber \\
    &=  \iint \dint{y_1} \dint{y_2} \dbar{\tw} 
    \left( \partial_{y_1} \trans{\start,y_1,\omega} \right) \trans{y_1,\target,\omega-\tw} \frac{\gamma (-\kappa)}{1 + \gamma \kappa \return{\target,\omega-\tw}} \left(\partial_{y_2} \trans{\target,y_2,\omega-\tw}\right) \trans{y_2,\target,\omega} \corr{\tw}
    \label{appeq:overlapping_diagram_2}
\end{align}
Inserting the result of \Eqref{appeq:overlapping_diagram_2} into \Eqref{appeq:QIII_final_loop}, one obtains
\begin{multline}\label{appeq:Q3_as_integral}
     \traceIII(\start, \target, \omega) \\
    =- \frac{1}{-i \omega} \frac{1}{\return{\target,\omega}} \iint \dint{y_1} \dint{y_2} \dbar{\tw} \frac{\left( \partial_{y_1} \trans{\start,y_1,\omega} \right) \trans{y_1,\target,\omega-\tw}}{\return{\target,\omega-\tw}} \left(\partial_{y_2} \trans{\target,y_2,\omega-\tw}\right) \trans{y_2,\target,\omega} \corr{\tw}
\end{multline}

Finally, we consider 
$\traceIV$, shown as IV) in Fig.~\ref{fig:four_diagrams}, where two $\cpl$-vertices couple  into the incoming transition propagator,
\begin{align}
\traceIV(\start,\target,\omega) &= \carryingCap^{-1} \lim_{\gamma \to \infty} \left[ \gamma \tikz[baseline=-5pt]{
    \draw[Aactivity] (0,0) -- (1.,0) ;
    \draw[Bsubstrate] (-.5,0) -- (0,0);
    \draw[Bactivity] (.75,0) arc (0:180:.25);
	\draw[very thick] (.85,-0.1) -- (0.85,0.1);
	\draw[very thick] (.35,-0.1) -- (.35,0.1);
   }
   -
    \gamma^2 \tikz[baseline=-5pt]{
    \draw[Aactivity] (-.5,0) -- (1.,0) ;
    \draw[Bsubstrate] (-1,0) -- (-.5,0);
    \draw[Bsubstrate] (-.5,0) arc(180:360:.25);
    \draw[Bactivity] (.75,0) arc (0:180:.25);
	\draw[very thick] (.85,-0.1) -- (0.85,0.1);
	\draw[very thick] (.35,-0.1) -- (.35,0.1);
   }
   +
    \gamma^3 \tikz[baseline=-5pt]{
    \draw[Aactivity] (-1,0) -- (1.,0) ;
    \draw[Bsubstrate] (-1.5,0) -- (-1.,0);
    \draw[Bsubstrate] (-.5,0) arc(180:360:.25);
    \draw[Bsubstrate] (-1.,0) arc(180:360:.25);
    \draw[Bactivity] (.75,0) arc (0:180:.25);
	\draw[very thick] (.85,-0.1) -- (0.85,0.1);
	\draw[very thick] (.35,-0.1) -- (.35,0.1);
   } + ... \right] \\
   &= \carryingCap^{-1} \lim_{\gamma \to \infty} \left[ \gamma \tikz[baseline=-5pt]{
    \draw[Bsubstrate] (-.5,0) -- (0,0);
    }\sum_{r=0}^{\infty} \left( \gamma \tikz[baseline=-5pt]{
    \draw[Aactivity] (-.5,0) -- (0,0) ;
    \draw[Bsubstrate] (-.5,0) arc(180:360:.25);
   } \right)^r
   \tikz[baseline=-5pt]{
    \draw[Aactivity] (0,0) -- (1,0);
	\draw[Bactivity] (.75,0) arc (0:180:.25);
	\draw[very thick] (.85,-0.1) -- (0.85,0.1);
	\draw[very thick] (.35,-0.1) -- (.35,0.1);    }  \right] \\
   &= \carryingCap^{-1} \lim_{\gamma \to \infty} \left[ \gamma \frac{1}{-i \omega} \frac{(-\lambda) \gamma}{1 + \gamma \kappa R(\target, \omega) } \right. \\
    & \qquad \qquad \qquad  \left.
   \times \iint \dint{y_1} \dint{y_2} \dbar{\tw} \left( \partial_{y_1} \trans{\start,y_1,\omega} \right) \left(\partial_{y_2} \trans{y_1,y_2,\omega-\tw}\right) \trans{y_2,\target,\omega} \corr{\tw}  \right] \nonumber  \\
     &= \label{appeq:Q4_as_integral}
     \frac{1}{-i \omega} \frac{1}{\return{\target,\omega}} \iint \dint{y_1} \dint{y_2} \dbar{\tw} \left( \partial_{y_1} \trans{\start,y_1,\omega} \right) \left(\partial_{y_2} \trans{y_1,y_2,\omega-\tw}\right) \trans{y_2,\target,\omega} \corr{\tw}
    \end{align}

The trace function corrected to leading order in the external noise is thus given by    
    \begin{align}
    \trace(\start,\target,\omega) = \frac{\trans{\start,\target,\omega}}{(-i\omega)\return{\target,\omega}} + \cpl^2 \left[\traceI + \traceII + \traceIII + \traceIV \right]
    +\cO(\cpl^3)
    \label{appeq:q_as_sum_of_4}
    \end{align}
Comparing the four correction terms, $\traceI,\ldots,\traceIV$, it turns out that they draw on two different integrals,
\begin{align}
\cJ_1(\start,\target,\omega) &=\iint \dint{y_1} \dint{y_2} \dbar{\tw} \left( \partial_{y_1} \trans{\start,y_1,\omega} \right) \left(\partial_{y_2} \trans{y_1,y_2,\omega-\tw}\right) \trans{y_2,\target,\omega} \corr{\tw} \\
\cJ_2(\start,\target,\omega)&=\iint \dint{y_1} \dint{y_2} \dbar{\tw} \frac{\left( \partial_{y_1} \trans{\start,y_1,\omega} \right) \trans{y_1,\target,\omega-\tw}}{\return{\target,\omega-\tw}} \left(\partial_{y_2} \trans{\target,y_2,\omega-\tw}\right) \trans{y_2,\target,\omega} \corr{\tw}
\end{align}
with $\cJ_1$ entering into $\traceI$ and $\traceIV$, Eqs.~\eqref{appeq:Q1_as_integral} and \eqref{appeq:Q4_as_integral}, and $\cJ_2$ entering into $\traceII$ and $\traceIII$, Eqs.~\eqref{appeq:Q2_as_integral} and \eqref{appeq:Q3_as_integral},
\begin{align}
\traceI(\start,\target,\omega) &= - \frac{1}{-i\omega} \frac{\trans{\start, \target, \omega}}{\left( \return{\target, \omega} \right)^2} \cJ_1(\target, \target, \omega) \\
\traceII(\start,\target,\omega) &= \frac{1}{-i\omega} \frac{\trans{\start, \target, \omega}}{\left( \return{\target, \omega} \right)^2} \cJ_2(\target, \target, \omega)  \\
\traceIII(\start,\target,\omega) &= - \frac{1}{-i\omega} \frac{1}{ \return{\target, \omega} } \cJ_2(\start, \target, \omega) \\
\traceIV(\start,\target,\omega) &= \frac{1}{-i\omega} \frac{1}{ \return{\target, \omega} } \cJ_1(\start, \target, \omega).
\end{align}
\Eqref{appeq:q_as_sum_of_4} simplifies further when factorising out the term to order $\cpl^0$:
\begin{align}
\trace(\start,\target,\omega) = \frac{\trans{\start,\target,\omega}}{(-i\omega)\return{\target,\omega}}\left[1 + \cpl^2 \left( \frac{\cJ_1(\start,\target,\omega) - \cJ_2(\start,\target,\omega)}{\trans{\start,\target,\omega}}  - \frac{\cJ_1(\target,\target,\omega) - \cJ_2(\target,\target,\omega)}{\return{\target,\omega}} \right) \right] +\cO(\cpl^3)
\label{appeq:trace_prob_result_1}
\end{align}
To simplify notation, we  introduce 
\begin{align}
\label{appeq:def_cal_K}
\transeff{\start,\target,\omega} &= \cJ_1(\start,\target,\omega) - \cJ_2(\start,\target,\omega) \\
&= \iint \dint{y_1} \dint{y_2} \dbar{\tw} \left[\left(\partial_{y_2} \trans{y_1,y_2,\omega-\tw}\right)
- \frac{ \trans{y_1,\target,\omega-\tw}}{\return{\target,\omega-\tw}}
\left(\partial_{y_2} \trans{\target,y_2,\omega-\tw}\right)
 \right] \\
 & \notag\qquad \qquad \qquad \qquad \times \left( \partial_{y_1} \trans{\start,y_1,\omega} \right) \trans{y_2,\target,\omega} \corr{\tw}\\
 &= \iint \dint{y_1} \dint{y_2} \label{appeq:T2_appendix}
 \left( \partial_{y_1} \trans{\start,y_1,\omega} \right) \left(\partial_{y_2} \trans{y_2,\target,\omega}\right)
  \\
 & \notag\qquad \qquad \qquad \qquad \times \int \dbar{\tw}  \left[ 
 \frac{ \trans{y_1,\target,\omega-\tw}}{\return{\target,\omega-\tw}}
 \trans{\target,y_2,\omega-\tw} - \trans{y_1,y_2,\omega-\tw}
 \right] \corr{\tw}
\end{align}
where the last equality follows by integration by parts and re-arranging terms, arriving at \Eqref{eq:T2_formula}.
Using $\transeff{\start,\target,\omega}$ in \Eqref{appeq:trace_prob_result_1} it may be written as
\begin{align}
\trace(\start,\target,\omega) &= \frac{T(\start,\target,\omega) + \cpl^2 \transeff{\start,\target,\omega}}{(-i\omega)\left( \return{\target,\omega} + \cpl^2\transeff{\target,\target,\omega} \right)} + \cO(\cpl^3) \ .
\end{align}
In keeping with the notation of $T^{(2)}$ as the $\cpl^2$-correction to the transition probability, we henceforth write $T^{(0)}$ for what used to be called $T$, the contribution at $\cpl=0$. Collecting these terms into the \emph{renormalised} $T$, we write
\begin{equation}
    T(\start,\target,\omega) = T^{(0)}(\start,\target,\omega) + \cpl^2 T^{(2)}(\start,\target,\omega) + \cO(\cpl^3)
\end{equation}
and along the same lines the return probability
\begin{equation}
R(\target,\omega) = T(\target,\target,\omega) 
    =R^{(0)}(\target,\omega) + \cpl^2 R^{(2)}(\target,\omega) + \cO(\cpl^3)
    =T^{(0)}(\target,\target,\omega) + \cpl^2 T^{(2)}(\target,\target,\omega) + \cO(\cpl^3) \ ,
\end{equation}
so that
\begin{equation}
    \trace(\start,\target,\omega) = \frac{T(\start,\target,\omega)}{(-i\omega)R(\target,\omega)} + \cO(\cpl^3) \ .
\end{equation}

\section{Derivation of effective transition probability for Brownian Motion driven by self-correlated noise \label{app:T2_Brownian_Motion}}
To compute $\transeff{x,\omega}$ in \Eqref{appeq:T2_appendix},
we firstly express the Fourier-transformed correlation function $\fcorr{\omega}$ in terms of the inverse Laplace transform $\invcorr{\beta}$, using
\begin{align}
	\fcorr{\omega} &= \int_{-\infty}^{\infty} \dint{t} e^{i \omega t} C(|t|) = \int_{-\infty}^{\infty} \dint{t} e^{i \omega t} \int_{0}^{\infty} \dint{\beta} e^{-\beta |t| } \invcorr{\beta} \\
	&=2 \int_{0}^{\infty} \dint{\beta} \left( \int_{0}^{\infty} \dint{t} \cos(\omega t) e^{-\beta t} \right) \invcorr{\beta} \\
	&= \int_{0}^{\infty} \dint{\beta} \frac{2\beta}{\omega^2 + \beta^2} \invcorr{\beta} \ ,
\end{align}
which facilitates the calculation of the convolution over $\tw$ in the second line of \Eqref{appeq:T2_appendix}, in particular when we consider exponential correlations, \Eqref{eq:Upsilon_exponential}, in which case $\invcorr{\beta}\propto\delta(\beta-\beta^*)$. We first consider the convolution of $C_2$ with $T^{(0)}$,
\begin{align}
	\int \dbar{\omega}\, \transm{y_1,y_2,\omega-\tw} \corr{\omega} &=   
	\int_0^{\infty} \dint{\beta} \int \dbar{\omega} 
	\transm{y_1,y_2,\omega-\tw}
	\frac{2 \beta \invcorr{\beta} }{\tw^2+\beta^2} \\
	& = \int_{0}^{\infty} \dint{\beta} \invcorr{\beta} \transm{y_1,y_2,\omega + i \beta} \ ,
\end{align}
where we have used that $\transm{y_1,y_2,\omega}$ cannot have any poles in the upper half-plane, because its inverse Fourier transform  $\transm{y_1,y_2,\tau}$ must vanish for all $\tau<0$. If there were any poles in the upper half-plane, the auxiliary path that for $\tau<0$ must pass through the upper half plane would enclose them, producing $\transm{y_1,y_2,\tau}\ne0$.

Considering secondly the convolution of $C_2$ with the term of the form $T^{(0)} T^{(0)}/R^{(0)}$ in \Eqref{appeq:T2_appendix}, we similarly obtain
\begin{align}
   \int \dbar{\tw} \frac{ \transm{y_1,\target,\omega-\tw}}{\returnm{\target,\omega-\tw}}
   \transm{\target,y_2,\omega-\tw} \corr{\tw} &= \int \dbar{\tw} \fptmgf^{(0)}(y_1,x_1,\omega - \tw) \transm{x_1,y_2,\omega - \tw} \int_0^{\infty} \dint{\beta} \frac{2 \beta \invcorr{\beta}}{\tw^2 + \beta^2}\\
   &= \int_{0}^{\infty} \dint{\beta} \invcorr{\beta} \fptmgf^{(0)}(y_1,x_1,\omega +i \beta) \transm{x_1,y_2,\omega + i \beta} \\
   & =
   \int_{0}^{\infty} \dint{\beta} \invcorr{\beta} 
   \frac{\transm{y_1,\target,\omega+i\beta}}{\returnm{\target,\omega+i\beta}}
   \transm{x_1,y_2,\omega + i \beta}
\end{align}
where we made use of the Markovian formula, Eqs.~\eqref{appeq:Q_in_omega} and \eqref{eq:def_fptdist},
$\fptmgf^{(0)}(\omega)=\transm{\start,\target;\omega}/\returnm{\target;\omega}$, which is, like $\transm{y_1,y_2,\omega}$ above, a Fourier transform of a probability density that vanishes for all $\tau<0$ and thus has no poles in the upper half-plane.

Having performed the convolutions over $\tw$, turning them into easier integrals over $\beta$, what remains are the two spatial integrals over $y_1$ and $y_2$,
\begin{multline}\label{appeq:spatial_integrals_left}
	\transeff{\start,\target,\omega} 	= \int_0^{\infty} \dint{\beta} \invcorr{\beta} \iint \dint{y_1} \dint{y_2} \left[\frac{ \transm{y_1,\target,\omega+i\beta}}{\returnm{\target,\omega+i\beta}}
 \transm{\target,y_2,\omega+i\beta} - \transm{y_1,y_2,\omega + i \beta} \right] 
 \\
 \times
 \left(\partial_{y_1} \transm{\start,y_1,\omega}\right) \left(\partial_{y_2} \transm{y_2,\target,\omega}\right) 
\end{multline}
We proceed by calculating $\transeff{\start,\target,\omega}$ for the particular case of Brownian Motion, which has transition propagator
\begin{align}\label{appeq:Brownian_transm}
	\transm{\start,\target,\omega} = \int \dbar{k} \frac{e^{ik(\target-\start)}}{-i\omega + D_x k^2} = \frac{e^{-|\target-\start|\sqrt{\frac{-i\omega}{D_x}}}}{\sqrt{-4i\omega D_x}} \ .
\end{align} 
Beginning with the simpler integrand in \Eqref{appeq:spatial_integrals_left}, the first term we consider is 
\begin{align}
	\cI_1(\start,\target,\omega)&=  \int_0^{\infty} \dint{\beta} \invcorr{\beta} \iint \dint{y_1} \dint{y_2}  \transm{y_1,y_2,\omega + i \beta}  \left(\partial_{y_1} \transm{\start,y_1,\omega}\right) \left(\partial_{y_2} \trans{y_2,\target,\omega}\right) \\
  &=  \int_0^{\infty} \dint{\beta} \invcorr{\beta} \iint \dint{y_1} \dint{y_2} \iint \dint{k} \dint{p} \dint{q} \frac{e^{ik(y_2-y_1)}}{-i\omega + D_x k^2+\beta}\frac{(ip) e^{ip(y_1-x_0)}}{-i\omega +  D_x p^2} \frac{(-iq) e^{iq(\target-y_1)}}{-i\omega +  D_x q^2}
\end{align}
Integration over both $y_1$ and $y_2$ results in two delta functions $\deltabar(k-p)$ and $\deltabar(p-q)$, respectively. Integrating over both $\dbar{k}$ and $\dbar{q}$ then results in
\begin{align}
	\cI_1(\start,\target,\omega) =  \int_{0}^{\infty} \dint{\beta} \invcorr{\beta} \int \dbar{p} \frac{ p^2 e^{ip(x_1 - x_0)}}{\left(-i\omega + D_x p^2\right)^2\left(-i\omega + D_x p^2 + \beta\right)} \ ,
\end{align}
which using partial fractions can be expressed in terms of the Markovian transition densities (\emph{cf} \cite[Eq.~(125)]{Walter2021}),
\begin{align}
	\cI_1(\start,\target,\omega) &=  \int_{0}^{\infty} \dint{\beta} \invcorr{\beta} \int \dbar{p} \left( \frac{1}{\beta}\frac{p^2e^{ip(x_1-x_0)}}{\left(-i\omega  + D p^2\right)^2} - \frac{1}{\beta^2} \frac{p^2e^{ip(x_1-x_0)}}{-i\omega + D p^2} + \frac{1}{\beta^2} \frac{p^2e^{ip(x_1-x_0)}}{-i\omega + D p^2 + \beta}  \right) \\
	&= -\int_{0}^{\infty} \dint{\beta} \invcorr{\beta} \beta^{-2} (\partial_{x_1}^2) \left[ -i \partial_{\omega} \beta \transm{x_0,x_1,\omega} - \transm{x_0,x_1,\omega} + \transm{x_0,x_1,\omega+i\beta} \right]
	\ .
\end{align}
Expressing $\transm{\start,\target,\omega}$ as a Fourier-transform in $t$, further leads to
\begin{align}
	\cI_1(\start, \target, \omega) &= 
	-\int_{0}^{\infty} \dint{\beta} \invcorr{\beta} 
	\int \dint{t} e^{i\omega t} 
	\beta^{-2}
	\left[ \beta t - 1 + e^{-\beta t} \right] 
	(\partial_{x_1}^2) \transm{x_0,x_1,t} \\
	&= -\int_{0}^{\infty} \dint{\beta} \invcorr{\beta} \int_{}^{} \dint{t} e^{i\omega t} t^2 Y\left( \beta t \right) (\partial_{x_1}^2) \transm{\start,\target,t}
\end{align}
where we introduced the dimensionless scaling function
\begin{align}
	Y(z) = \frac{e^{-z}-1 + z}{z^2}
	\label{}
\end{align}
with $Y(z) \stackrel{z \to 0}{\rightarrow} 1/2$ and $Y(z) \stackrel{z \gg 1}{\simeq} z^{-1}$.
While $Y(z)$ is specific to Brownian Motion, other stochastic processes give rise to similar scaling functions as \cite[Eq.~(97)]{Walter2021} indicates for the case of an Ornstein-Uhlenbeck process.

The second contribution of \Eqref{appeq:spatial_integrals_left} is 
\begin{align}
  &\cI_2(\start,\target,\omega) \\
  &= \int_{0}^{\infty} \dint{\beta} \invcorr{\beta} \iint \dint{y_1} \dint{y_2} \frac{ \transm{y_1,\target,\omega+i\beta}}{\returnm{\target,\omega+i\beta}}
 \transm{\target,y_2,\omega+i\beta}  \left(\partial_{y_1} \trans{\start,y_1,\omega}\right) \left(\partial_{y_2} \transm{y_2,\target,\omega}\right) \\
 &=\int_{0}^{\infty} \dint{\beta} \invcorr{\beta}
 \sqrt{4D_x(\beta-i\omega)}
 \iint \dint{y_1} \dint{y_2} \iint \dbar{k_1}\dbar{k_2}\dbar{p}\dbar{q} \\
 & \nonumber\qquad\qquad\qquad
 \frac{e^{ik_1(x_1-y_1)}}{-i\omega +  D_x k_1^2+\beta}
  \frac{e^{ik_2(y_2-x_1)}}{-i\omega + D_x k_2^2 + \beta}
 \frac{(ip)e^{ip(y_1-\start)}}{-i\omega  + D_x p^2} \frac{(-iq)e^{iq(\target-y_2)}}{-i\omega  + D_x q^2} \ ,
\end{align}
using $1/\returnm{\target,\omega+i\beta}=1/\transm{\target,\target,\omega+i\beta}=\sqrt{4D_x(\beta-i\omega)}$, \Eqref{appeq:Brownian_transm}.
Integrating over $y_1$ and $y_2$ produces two $\delta$-functions,  $\delta(p-k_1)\delta(q-k_2)$. Using them when integrating over $k_1, k_2$ gives
\begin{multline}
	\cI_2(\start,\target,\omega)  = \int_{0}^{\infty} \dint{\beta} \invcorr{\beta} \sqrt{4D_x(\beta-i\omega)} \\
   \times \left(\int \dint{q} \frac{(-iq)}{(-i\omega +D_x q^2)(-i\omega  +D_x q^2 + \beta)} \right) 
     \left( \int \dint{p} \frac{(ip)e^{ip(x_1-x_0)}}{(-i\omega +D_x p^2)(-i\omega  +D_x p^2 + \beta)} \right)
\end{multline}
By symmetry, the integral over $\dbar{q}$ vanishes and thus $\cI_2 = 0$. 
For Brownian Motion, we thus obtain for $\transeff{\start,\target,\omega}$, \Eqrefs{eq:T2_formula} and \eqref{appeq:T2_appendix},
\begin{align}
	\transeff{\start,\target,\omega} = -\cI_1(\start,\target,\omega) = \int_{0}^{\infty} \dint{\beta} \invcorr{\beta} \int_{}^{} \dint{t} e^{i\omega t} t^2 Y\left( \beta t \right) (\partial_{x_1}^2) \transm{\start,\target,t} 
\end{align}
By inverting the Fourier transform we further obtain
\begin{align}
	\transeff{\start,\target,t} &= \int_{0}^{\infty} \dint{\beta} \invcorr{\beta}  t^2 Y\left( \beta t \right) (\partial_{x_1}^2) \transm{\start,\target,t} \\
	&= \Upsilon(t) (\partial_{x_1}^2) \transm{x_0,x_1,t}
\end{align}
where we introduced the time-stretch function, \Eqref{eq:def_upsilon},
\begin{align}
	\Upsilon(t) &= \int_{0}^{\infty} \dint{\beta} \invcorr{\beta} t^2 Y(\beta t) \\
	&= \int_{0}^{\infty} \dint{\beta} \frac{\invcorr{\beta}}{\beta^2} \left[ e^{-\beta t} - 1 + \beta t \right] 
	\ .
\end{align}
Making use of the properties of the Laplace transform, we find
\begin{align}
	\int_{0}^{\infty} \dint{\beta} \beta^{-2} e^{-\beta t} \invcorr{\beta} &= \int_{0}^{\infty} \dint{s} s \corr{t + s} \\
	\int_{0}^{\infty} \dint{\beta} \beta^{-2} \invcorr{\beta} &= \int_{0}^{\infty} \dint{s} s \corr{s} \\
	\int_{0}^{\infty} \dint{\beta} \beta^{-1}t  \invcorr{\beta} &=  \int_{0}^{\infty} \dint{s} t \corr{ s} 
\end{align}
and after some simple transformations,
\begin{align}
	\Upsilon(t) &= 
	\int_0^\infty \dint{s} \Big(s \corr{t + s} - s \corr{s}  + t \corr{s} \Big)
	= \int_{0}^{t} \dint{s} (t-s) \corr{s} 
	=\int_0^t \dint{s} \int_0^s \dint{u} \corr{u}\ , 
\end{align}
as in \Eqref{eq:def_upsilon}.

\section{List of explicit results for visit probabilities \label{app:list_of_results}}
The concrete perturbative corrections resulting from formulas \eqref{eq:nonM_trace} and \eqref{eq:T2_formula} are often cumbersome expressions. The correction for the active thermal Brownian Motion on a real line, characterised by \eqref{eq:ATBM_Langevin}, and with an exponentially correlated driving noise is given implictly via \Eqref{eq:nonM_trace_brownian}.
This integral can be performed using Mathematica \cite{Mathematica} and delivers
\begin{align}
 \label{appeq:Q_line}
    \trace(x,\omega) = \frac{e^{-\sqrt{\frac{-i\omega x^2}{D_x}}}}{-i\omega} \left[1 + \frac{\cpl^2 D_y}{D_x^2 \beta} \left( 
   D_x \sqrt{(-i\omega)(\beta - i \omega)} \left(e^{- \frac{\sqrt{\beta - i \omega} - \sqrt{-i\omega}}{\sqrt{D_x}} |x|} -1 \right)
   + \frac12 \beta \sqrt{-i D_x \omega} |x|
    \right)   \right]
\end{align}
In the joint limit of $\beta \to 0$ and $D_y \beta = w^2$ fixed, the moment generating function becomes
\begin{align}
    \trace(x,\omega) = \frac{e^{-\sqrt{\frac{-i\omega x^2}{D_x}}}}{-i\omega} \left[1 + \frac{\cpl^2 w^2}{8 D_x^2} \left( 
    x^2 - \sqrt{\frac{D_x x^2}{-i\omega}}
    \right)   \right]
\end{align}
This corresponds to the visit probability of a Brownian motion with a random but fixed additional drift term $y$ which is Gaussian distributed with mean zero and variance $w^2$.

In \cite{Walter2021}, we developed a perturbative framework which was able to compute the moment-generating functions of first-passage time distributions for processes of the form \eqref{eq:driven_langevin}. This framework did not use field theory, but instead a functional perturbation theory. Since $\trace(x,\omega) = \frac{1}{-i \omega} \fptmgf(x,\omega)$, we here report the findings for two other models first reported there, for future reference.

First, we report the visit probability of an active Brownian Motion on a ring of radius $r$, hence
\begin{align}
    \dot{x}_t = \xi_t + \cpl y_t \qquad x_t \equiv x_t + 2 \pi r.
\end{align}
We then study the visit probability over a certain angle $\theta = \frac{x_1 - x_0}{r}$.
As a shorthand, we further introduce the inverse diffusive timescale $\alpha^{-1} = r^2/D_x$. To leading perturbative order, the visit probability is then given by \cite[Eq.~(127)]{Walter2021}
\begin{align}
\label{appeq:Q_ring}
    \trace(\theta,\omega) &=\frac{\cosh\left( (\theta - \pi)\sqrt{ - i \alpha^{-1} \omega } \right)}{ (-i\omega) \cosh\left( \pi\sqrt{- i \alpha^{-1} \omega}  \right)} 
	\\
	&+ \frac{D_y \cpl^2}{2 D_x} \cdot
	\frac{\sqrt{- i \alpha^{-1} \omega}\tanh\left( \pi \sqrt{- i \alpha^{-1} \omega} \right)}{-i \alpha^{-1} \beta \omega} \left[ \frac{\cosh\left( (\theta - \pi)\sqrt{\alpha^{-1}(\beta - i \omega)} \right)}{\sinh\left( \pi\sqrt{\alpha^{-1}(\beta - i \omega)} \right)}2\sqrt{\alpha^{-1}(\beta - i \omega)}\right.
		\nonumber \\
	&  + \left.\frac{\cosh\left( (\theta-\pi) \sqrt{- i \alpha^{-1} \omega} \right)}{ \cosh\left( \pi \sqrt{- i \alpha^{-1} \omega} \right)} \left( \pi\bar{\beta}-2\sqrt{\alpha^{-1}(\beta - i \omega)} \coth\left( \pi \sqrt{\alpha^{-1}(\beta - i \omega)} \right) \right)+ \frac{\sinh\left( (\theta-\pi)\sqrt{- i \alpha^{-1} \omega} \right)}{\sinh\left( \pi\sqrt{- i \alpha^{-1} \omega} \right)}\bar{\beta}\left( \pi-\theta \right) \right] 
	\nonumber
\end{align}
In the limit of infinite radius, $r \to \infty$, one finds $\theta \sqrt{-i \alpha^{-1} \omega} \to \sqrt{-i \frac{\omega(x_1 - x_0)^2}{D_x}}$, and accordingly the Markovian result converges, as expected to
\begin{align}
    \lim_{r \to \infty} \frac{\cosh\left( (\theta - \pi)\sqrt{ - i \alpha^{-1} \omega } \right)}{ (-i\omega) \cosh\left( \pi\sqrt{- i \alpha^{-1} \omega}  \right)} = \lim_{r \to \infty}  \frac{1}{(-i\omega)}\frac{\cosh\left( \sqrt{-i \frac{\omega}{D_x}} |x_1 - x_0| - \sqrt{-i \frac{\omega}{D_x}} \pi r \right)}{\cosh( \sqrt{-i \frac{\omega}{D_x}} \pi r )} = \frac{e^{\sqrt{-i \frac{\omega}{D_x}} |x_1 - x_0|}}{-i\omega}
\end{align}
Analagously, a more involved computation using, for instance, Mathematica confirms that
\begin{align}
    \lim_{r \to \infty} \trace^{\text{ring}}( \frac{x_1 - x_0}{r}, \omega) = \trace^{\text{line}}( x_1 - x_0, \omega)
\end{align}
with $\trace^{\text{ring}}( \frac{x_1 - x_0}{r}, \omega)$ being the result in \Eqref{appeq:Q_ring} and $\trace^{\text{line}}( x_1 - x_0, \omega)$ the perturbative result found in \Eqref{appeq:Q_line}.

Finally we consider the case of a harmonic trap, \ie 
\begin{align}
    \dot{x}_t = - \alpha x + \xi_t + \cpl y_t.
\end{align}
which we refer to as active thermal Ornstein Uhlenbeck process (ATOU). The result is more compactly given in dimensionless units
\begin{align}
    \bar{\beta} = \alpha^{-1} \beta \qquad \bar{x}_0 = x_0/\ell \qquad \bar{x}_1 = x_1/ \ell \qquad \text{with} \qquad \ell = \sqrt{D_x \alpha^{-1}} 
\end{align}
The Fourier transformed visit probability is given in terms of parabolic cylinder functions $D_{-\nu}(x)$ (\cite{gradshteyn_table_2007}) and reads (for $x_0 < x_1)$ \cite[Eq.~(100)]{Walter2021}
\begin{align}
    \trace(x,\omega) =& e^{\frac{\bar{x}_0^2 - \bar{x}_1^2 }{4}} \frac{D_{i \alpha^{-1} \omega}(\bar{x}_0)}{(-i\omega)D_{i \alpha^{-1} \omega}(\bar{x}_1)}\\ 
   &
+ \frac{g^2 D_y \bar{\beta}}{D_x(-i\omega)} \frac{(-i \alpha^{-1} \omega) e^{\frac{{\bar{x}_0}^2-{\bar{x}_1}^2}{4}}}{2 \left({\bar{\beta}} ^2-1\right) D_{i \alpha^{-1} \omega }(-\bar{x}_1){}^2 D_{-{\bar{\beta}} + i \alpha^{-1} \omega }(-\bar{x}_1)}
\nonumber\\
&\times \left[ \left({\bar{\beta}} +1\right)(-i \alpha^{-1} \omega + 1)D_{-{\bar{\beta}}+i \alpha^{-1} \omega}(-\bar{x}_1) \left(D_{i \alpha^{-1} \omega}(-\bar{x}_0)D_{i \alpha^{-1} \omega-2}(-\bar{x}_1)-D_{i \alpha^{-1} \omega-2}(-\bar{x}_0)D_{i \alpha^{-1} \omega}(-\bar{x}_1) \right)\right. 
\nonumber\\
 & \qquad \left. -2 ({\bar{\beta}} - i \alpha^{-1} \omega) D_{i \alpha^{-1} \omega -1}(-\bar{x}_1) \left(D_{i \alpha^{-1} \omega }(-\bar{x}_0) D_{-{\bar{\beta}}+ i \alpha^{-1} \omega -1}(-\bar{x}_1)- D_{-{\bar{\beta}} + i \alpha^{-1} \omega -1}(-\bar{x}_0)D_{i \alpha^{-1} \omega }(-\bar{x}_1)\right)\right]
 \nonumber
\end{align}

\end{widetext}
\end{document}